\documentclass{aa}
\usepackage{graphicx}
\usepackage{txfonts}
\usepackage[normalem]{ulem}
\usepackage{xcolor}
\usepackage{subcaption}

\usepackage{lscape}
\usepackage{placeins}

\usepackage[colorlinks=true,allcolors=blue]{hyperref}
\usepackage{multirow}

\newcommand{\HI}{H\,{\sc i}}

\begin{document}

\title{A MUSE and MeerKAT study of dynamics and star formation in HCG 16}

\author{I. Labadie-García\fnmsep\thanks{Corresponding author: ixakalab@iaa.es}
    \and E. Pérez \and J. Garrido \and R. Ianjamasimanana \and L. Verdes-Montenegro}

\institute{Instituto de Astrofísica de Andalucía, Glorieta de la Astronomía s/n, 18008 Granada, Spain}

\date{\today}

\abstract{Hickson Compact Groups (HCGs) present an ideal environment for studying galaxy interactions. HCG 16 is a group of five galaxies in an intermediate stage of evolution (Phase 2), with several members displaying signs of interaction, including distorted morphologies, tidal features, and enhanced star formation rates.}
    {We aim to better understand the evolution of HCG 16 and its member galaxies, focusing on the relationship between neutral hydrogen and star formation, while reconstructing the interaction history of the group.}
    {We analysed MUSE data processed with pyPipe3D to extract H$\alpha$ images and stellar population synthesis results, and reduced MeerKAT data to produce a high-resolution \HI\ cube. We examined the structure and dynamics of the H$\alpha$ and \HI\ lines in both 2D and 3D, and produced spatially resolved star formation histories to compare the inner and outer parts of the galaxies.}
    {We identify a major star formation episode in HCG 16 beginning $\sim400$--500 Myr ago, coinciding with the passage of NGC 848 close to the other member galaxies, and that continued at least until 50--60 Myr ago in some galaxies. The \HI\ distribution is highly perturbed in the group, although HCG 16a retains a nearly regular disk, while HCG 16b has associated H$\alpha$ tails.}
    {The passage of NGC 848 probably activated star formation in HCG 16, but cannot by itself explain the current starburst activity in HCG 16c and 16d. The evolution of the group is better explained by multiple interactions among its members than by a single strong encounter.}

\maketitle
\nolinenumbers

\section{Introduction}

The environment in which galaxies reside plays a fundamental role in shaping their evolution \citep{dressler1980, postman1984, zandivarez2022}. Hickson Compact Groups (HCGs; \citealp{Hickson1982}) occupy a distinctive regime in which 4 to 10 galaxies are confined within small volumes and exhibit low velocity dispersions. These conditions favour frequent tidal interactions that redistribute gas and strongly influence star formation (SF). \cite{VerdesMontenegro2001} proposed an evolutionary scenario for HCGs based on the distribution of neutral hydrogen (\HI): HCGs evolve from an early stage (Phase 1), with most \HI\ confined to galaxy disks, through an intermediate phase (Phase 2), where tidal interactions redistribute gas into tidal features and the intra-group medium (IGrM), to a late stage (Phase 3), where most or all \HI\ is stripped from galaxy disks and, if present, resides in tidal tails. \cite{jonas2023} made minor modifications to this sequence.

\begin{figure}
    \centering
    \includegraphics[width=0.45\textwidth]{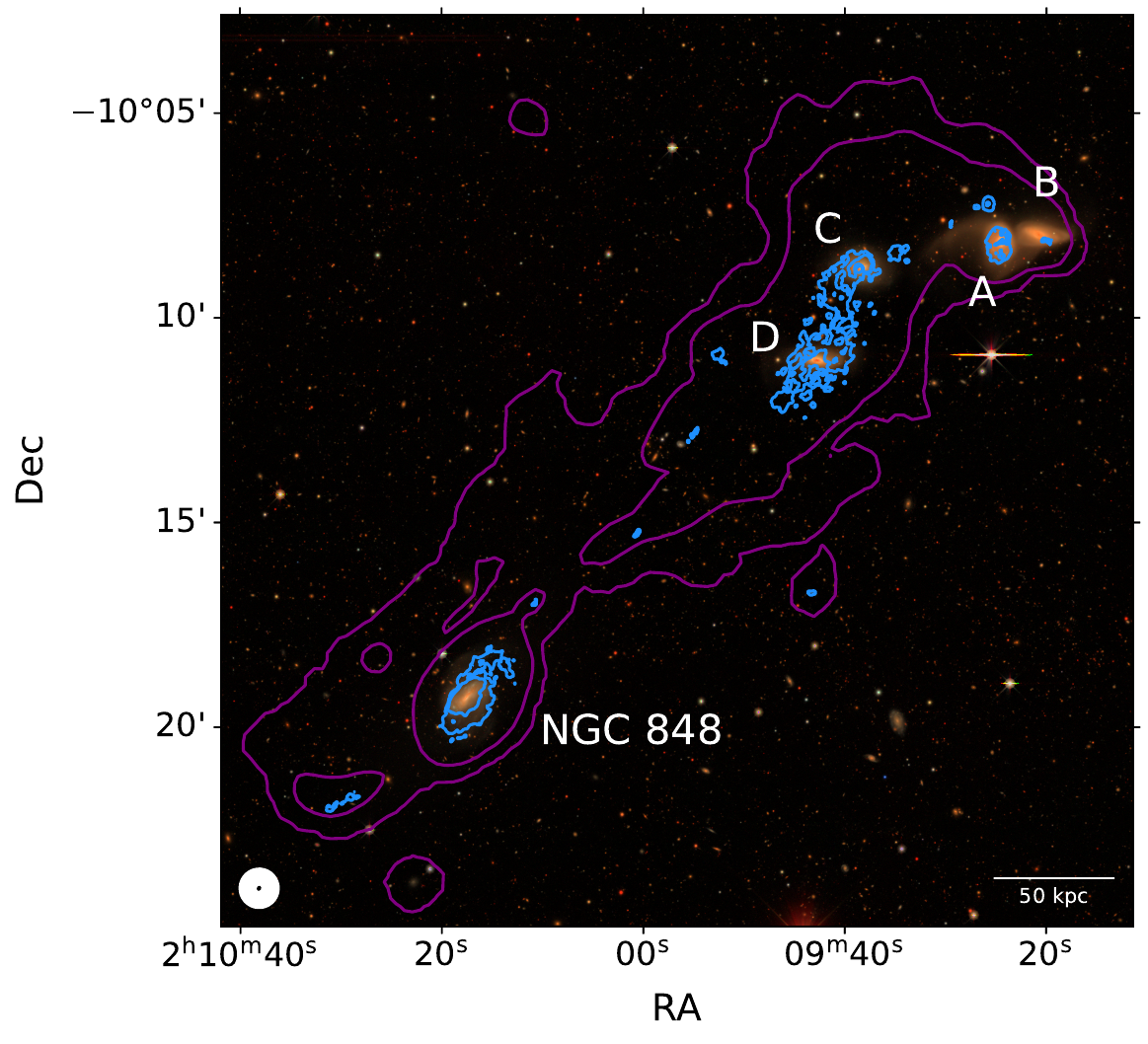}
    \caption{DECaLS DR10 \textit{rgz} image of HCG 16 overlaid with MeerKAT \HI\ column density contours from two different angular resolution cubes. Purple (outer) contours correspond to $3.1\times10^{18}$ cm$^{-2}$ ($3\sigma$ sensitivity) and $5\times10^{19}$ cm$^{-2}$ column density taken from low resolution data ($57.9''\times57.5''$ beam) presented by \cite{ianja2025}. Blue (inner) contours correspond to $1.5\times10^{20}$ cm$^{-2}$ ($3\sigma$ sensitivity) and $1.0\times10^{21}$ cm$^{-2}$ column density from high-resolution data ($9.8''\times6.7''$ beam) of this article. The white ellipse represents the low-resolution beam size and the black ellipse within, the high-resolution one.}
    \label{fig:decals}
\end{figure}

HCG 16, shown in Figure \ref{fig:decals}, is a typical Phase 2 group which was originally identified as a quartet \citep{Hickson1982}. That quartet is confined within 120 kpc and forms the core of the group. Subsequent studies expanded the group to seven members within 500 kpc \citep{deCarvalho1997,ribeiro1998}, although only NGC 848 was considered associated with the core. This was later confirmed by the detection of a 160 kpc \HI\ tail connecting NGC 848 to the core \citep{VerdesMontenegro2001}. More recently, \cite{ianja2025} increased the number of galaxies within the virial radius to nine, of which only five are believed to be interacting: HCG 16a (NGC 835), HCG 16b (NGC 833), HCG 16c (NGC 838), HCG 16d (NGC 839), and NGC 848.

While the total \HI\ content of the group is slightly higher than that of isolated galaxies with similar optical luminosity \citep{Jones2019,sorgho2025}, the gas is very redistributed: HCG 16a and 16b are significantly deficient whereas over $40\%$ of the total \HI\ lies outside galaxy disks \citep{sorgho2025}. This redistribution is thought to result from tidal stripping during a close passage of NGC 848 through the group core about 400--500 Myr ago, together with ongoing interactions among the core galaxies \citep{vogt2013,osullivan2014,Jones2019}.

HCG 16 has been widely studied in multiple wavelengths. \cite{mendesDeOliveira1998} analysed its morphology and kinematics using H$\alpha$ Fabry-Perot interferometry, while \cite{rich2010} and \cite{vogt2013} used optical integral field spectroscopy (IFS) to study galactic winds and SF in HCG 16d and 16c respectively. \cite{osullivan2014} conducted a multi-wavelength study of the group's nuclear activity, SF, and X-ray properties, using Chandra, VLA, GMRT, and SDSS data when available. \cite{Jones2019} examined the interactions and evolutionary phase of the group from its \HI\ content, morphology, and kinematics. They characterise most tidal features in HCG 16, including a tidal dwarf galaxy studied in detail by \cite{roman2021}. These structures were found to be more pronounced in MeerKAT observations \citep{ianja2025}.

The aim of this work is to better understand the evolution of HCG 16 by studying the relation between SF and neutral hydrogen, since the latter constitutes the primary reservoir from which molecular clouds form, ultimately providing the fuel for SF. We combine MUSE/VLT \citep{bacon2010} data with high-resolution MeerKAT \citep{jonas2016} observations to analyse the distribution and kinematics of H$\alpha$ and \HI, together with the novel, spatially resolved, star formation histories (SFH) of the group galaxies.

\section{Observations}

We used \HI\ data from a MeerKAT campaign that observed six HCGs in Phase 2 and Phase 3 of their evolutionary stages \citet{ianja2025}. In this paper, we use the high-resolution version of the HCG 16 data cube, which has a beam size of $9.8'' \times 7.0''$ and a spectral resolution of 5.7 km/s. The corresponding linear resolution is $\sim 2.3$ kpc at a distance of $58$ Mpc. Distances to the group and to individual galaxies have been estimated from the redshift given by NED\footnote{\url{https://ned.ipac.caltech.edu/}} with Planck18 cosmological parameters \citep{Planck2020}, except for HCG 16a, which was taken from SIMBAD \citep{simbad2000}. The data were processed with CARACal \citep{caracal2020} and SoFiA-2 \citep{sofia2015,sofia2021} to produce velocity cubes and moment maps, obtaining an $rms$ value of 0.31 mJy/beam and a $3\sigma$ column density sensitivity of $1.61\times 10^{20}$ cm$^{-2}$ assuming an \HI\ line width of $20$ km/s.

Figure \ref{fig:decals} shows the \HI\ extent of the cube presented by \cite{ianja2025}, which has a $57.9''\times57.5''$ beam, alongside that of the high-resolution cube used in this work. The high-resolution data reveal significantly less \HI, particularly in the IGrM. The total \HI\ mass detected in the high-resolution cube is $10.5\times10^{9}$ M$\odot$, compared to $23.9\times10^{9}$ M$\odot$ in the low-resolution cube. Figure \ref{fig:int_spect} compares the integrated spectra of the two cubes, following Figure 4 of \cite{ianja2025}. The high-resolution cube shows lower integrated \HI\ emission and covers a slightly narrower velocity range, although the difference is more pronounced for some individual galaxies.

\begin{figure}
    \centering
    \includegraphics[width=0.5\textwidth]{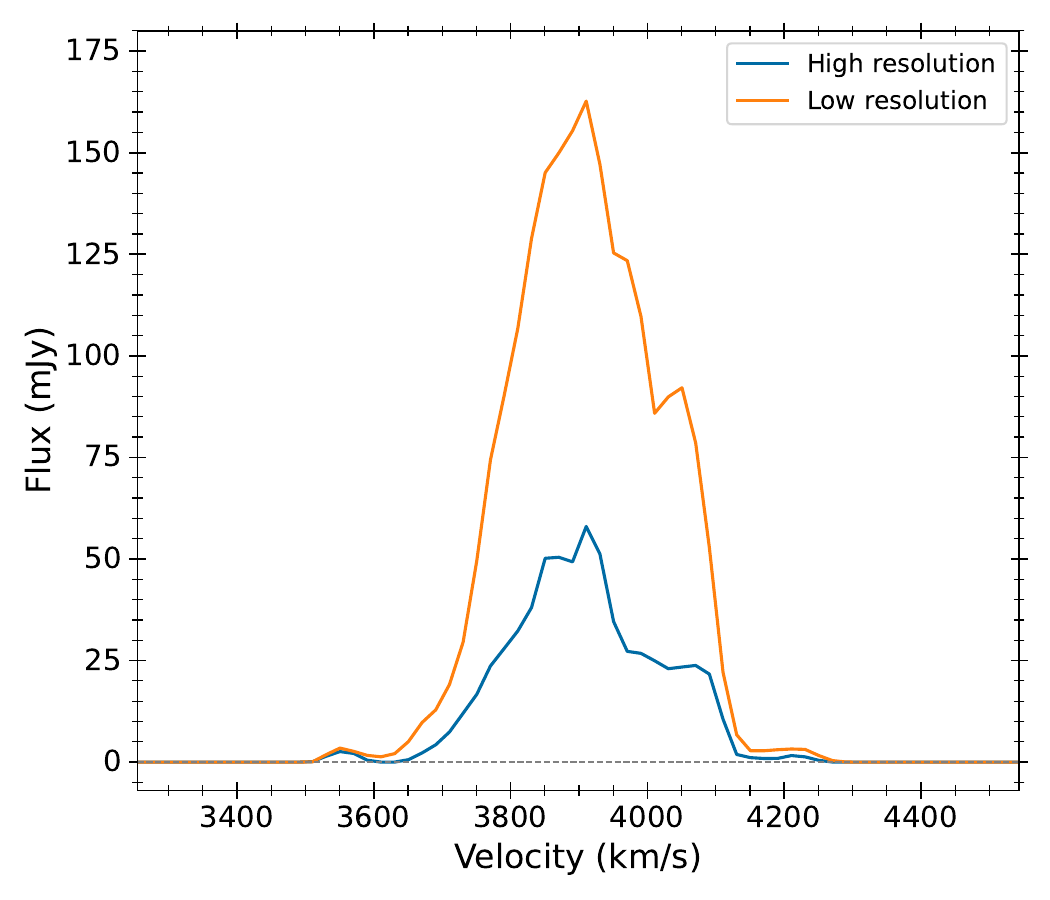}
    \caption{Integrated spectra from the high-resolution (blue) and low-resolution (orange) datacubes extracted from the area within their corresponding SoFiA masks --- areas with reliable \HI\ emission ---, smoothed at 20 Jy km/s.}
    \label{fig:int_spect}
\end{figure}

MUSE archive data are available for the five galaxies in HCG 16\footnote{ID: 096.B-0951, 097.B-0518; Joanna Woo.}, taken in Wide Field Mode (WFM). The emission line fitting and stellar population synthesis (SPS) performed with pyPipe3D\footnote{\url{https://ifs.astroscu.unam.mx/pyPipe3D/}} \citep{sanchez2016_pipe3d,lacerda2022_pyfit3d} are publicly available as part of the AMUSING++ compilation \citep{lopezCoba2020_amusing1}. They use the gsd156 \citep{cidFernandes2013} simple stellar population (SSP) library, and produce files with spectral line information, parameters of the SPS, light fraction for age and metallicity bins, and others\footnote{pyPipe3D products for MUSE: \url{https://ifs.astroscu.unam.mx/MUSE/dataproducts_NEW/}}. Cubes containing only emission lines are created as a middle step in pyPipe3D and were provided by Sebastian F. Sánchez.

In WFM, MUSE has a field of view of $1'\times1'$ with an effective spatial resolution of $\sim1''$ (FWHM), and a spectral range from 4750 to 9352 \AA{} with a resolution of 2.4 \AA{} (FWHM) at $\sim 7500$ \AA, corresponding to $\sim 100$ km/s for H$\alpha$. Comparing MUSE with MeerKAT, the spatial resolution of MUSE is $\sim 7$ times higher while the spectral resolution is around $17$ times lower. Nevertheless, the general comparison we aim to make does not require the two datasets to be at the same resolution.

\section{Results and discussion}

\subsection{Star formation history}

\begin{figure*}
    \sidecaption
    \includegraphics[width=0.70\textwidth]{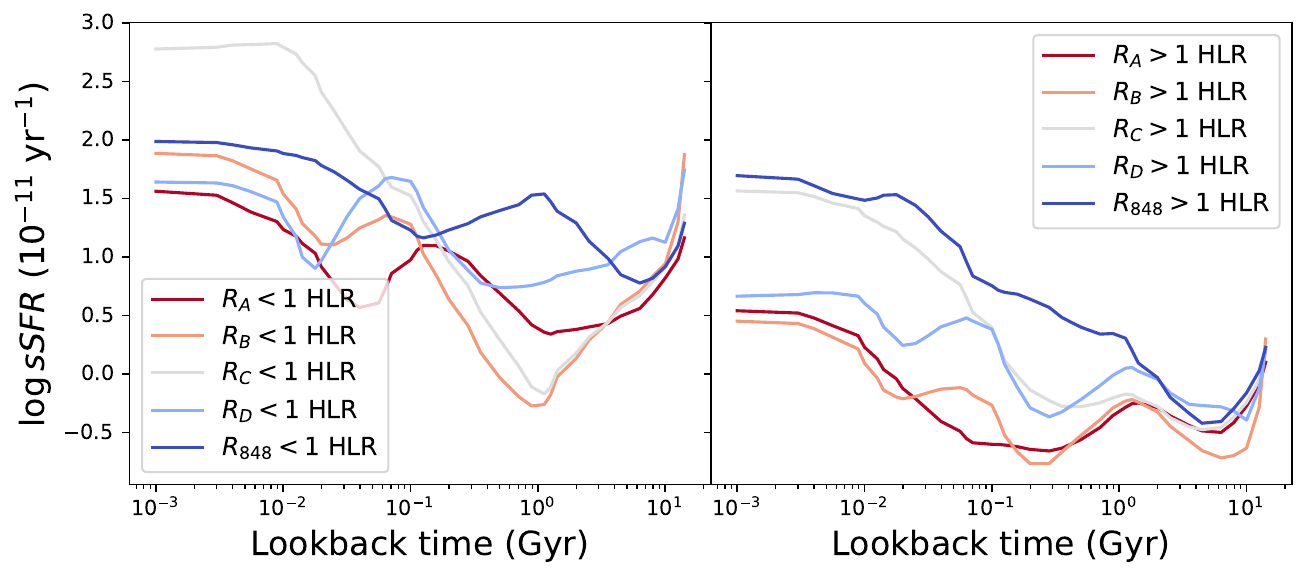}
    \caption{SFH inside (left) and outside (right) the HLR, convolved with a $\approx 0.25$ dex Gaussian in the lookback time axis. The sSFR is the SFR weighted by the total mass at each epoch. The galaxies are ordered from most massive (red) to least massive (blue).}
    \label{fig:hlr_sfh}
\end{figure*}

We computed the stellar mass formed at each epoch and the corresponding star formation rate (SFR) for each galaxy using the SSP and SFH output files from pyPipe3D, which contain the luminosity fraction per pixel at different ages (normalised to a reference flux intensity map at $\sim 5500$ \AA{}) and the corresponding mass-to-light ratio. The gsd156 library combines specific models for young and old stellar populations, with the transition set at 63 Myr. This produces a very narrow age bin (0.1 Myr) that contains a comparatively large light fraction given its width. Because such a narrow bin is unlikely to provide a reliable representation at that epoch, we excluded it from our analysis. To study the spatial behaviour, we fitted elliptical isophotes from the centre out to 4 half-light radii (HLR) and computed the SFR in radial bins inside and outside the HLR. To highlight global features and account for SPS uncertainties --- larger at small scales --- we applied a Gaussian filter of $\approx 0.25$ dex in the age axis and 0.2 HLR in the radial axis.

Figure \ref{fig:hlr_sfh} shows the specific SFR (sSFR) of each galaxy, defined as the SFR divided by the accumulated stellar mass at each epoch, inside and outside the HLR. We identify a peak at 1--2 Gyr in the outer regions of all five galaxies, and in the inner regions of some of them. This feature is common in most spiral galaxies \citep{Gonzalez2017} and is associated with post-AGB stars. We also find a peak at $\sim 60$ Myr, associated with B-type stars, in HCG 16b and 16d, while in HCG 16c the sSFR continues to increase beyond this time. Finally, we observe an increase at 10-12 Myr in all galaxies, which is sustained to the present day and is associated with O-type stars. Within the HLR, the post-AGB peak is clearly detected only in NGC 848 and marginally in HCG 16d. HCG 16b and 16c show the same behaviour, with a pronounced minimum at 1-2 Gyr followed by a strong increase, whereas HCG 16a is intermediate, with more A-type but fewer B-type stars than HCG 16b. This agrees with the UV flux ratios reported in \cite{tzanavaris2010}: HCG 16b and 16d are redder, with flux ratios exceeding a factor of two between the \textit{uvw1} (2634 \AA) and the \textit{uvw2} (2030 \AA) or \textit{uvm2} (2231 \AA) bands, while HCG 16a and 16c show lower values (1.2–1.5).

\begin{figure}[!ht]
    \centering
    \includegraphics[width=0.45\textwidth]{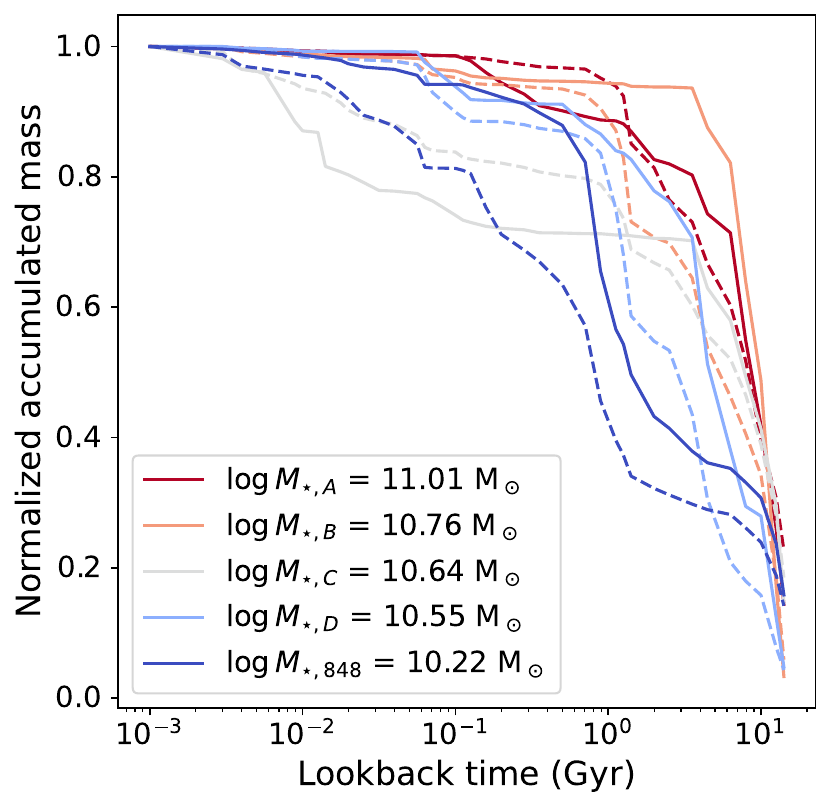}
    \caption{Normalised accumulated mass at different ages for all galaxies. Galaxies are ordered from most massive (red) to least massive (blue). Solid lines indicate regions within the HLR and dashed lines outside the HLR.}
    \label{fig:acc_mass}
\end{figure}

Figure \ref{fig:acc_mass} shows the cumulative stellar mass at different ages, separately for regions inside and outside the HLR and normalised to the present-day total stellar mass. As expected, more massive galaxies assembled most of their mass earlier, while lower-mass galaxies evolved more gradually. However, HCG 16c assembled large amounts of mass inside the HLR within the last 100 Myr. This is consistent with the increase seen in Figure \ref{fig:hlr_sfh}, where the sSFR of HCG 16c increases beyond 60 Myr until the present, unlike the flattening or decrease seen in other galaxies. Outside the HLR, NGC 848 also shows elevated recent sSFR, comparable to HCG 16c, possibly reflecting its slower secular evolution.

\begin{figure*}[!htbp]
    \centering
    \begin{subfigure}{0.49\textwidth}
        \includegraphics[width=\textwidth]{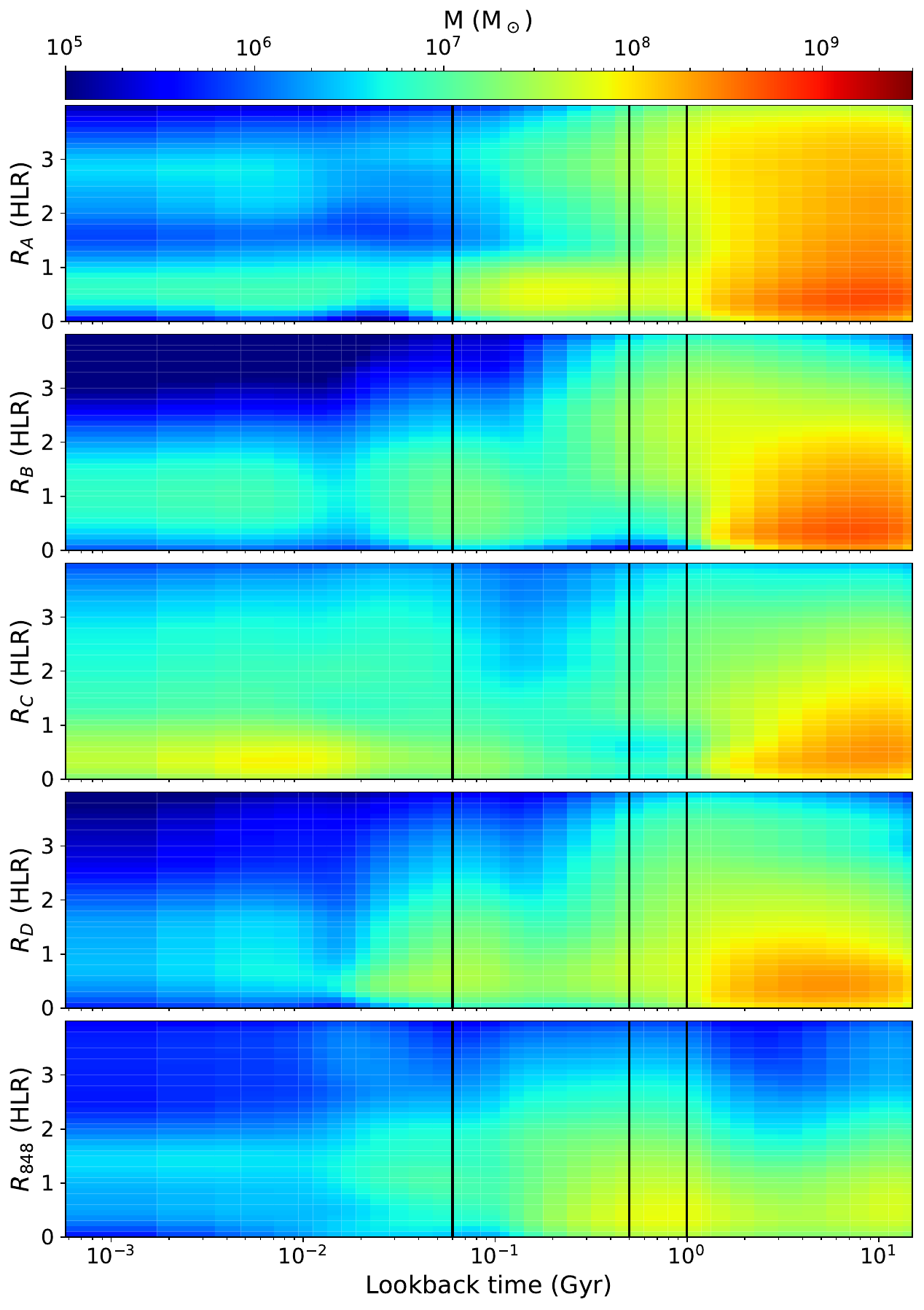}
    \end{subfigure}
    \hfill
    \begin{subfigure}{0.49\textwidth}
        \includegraphics[width=\textwidth]{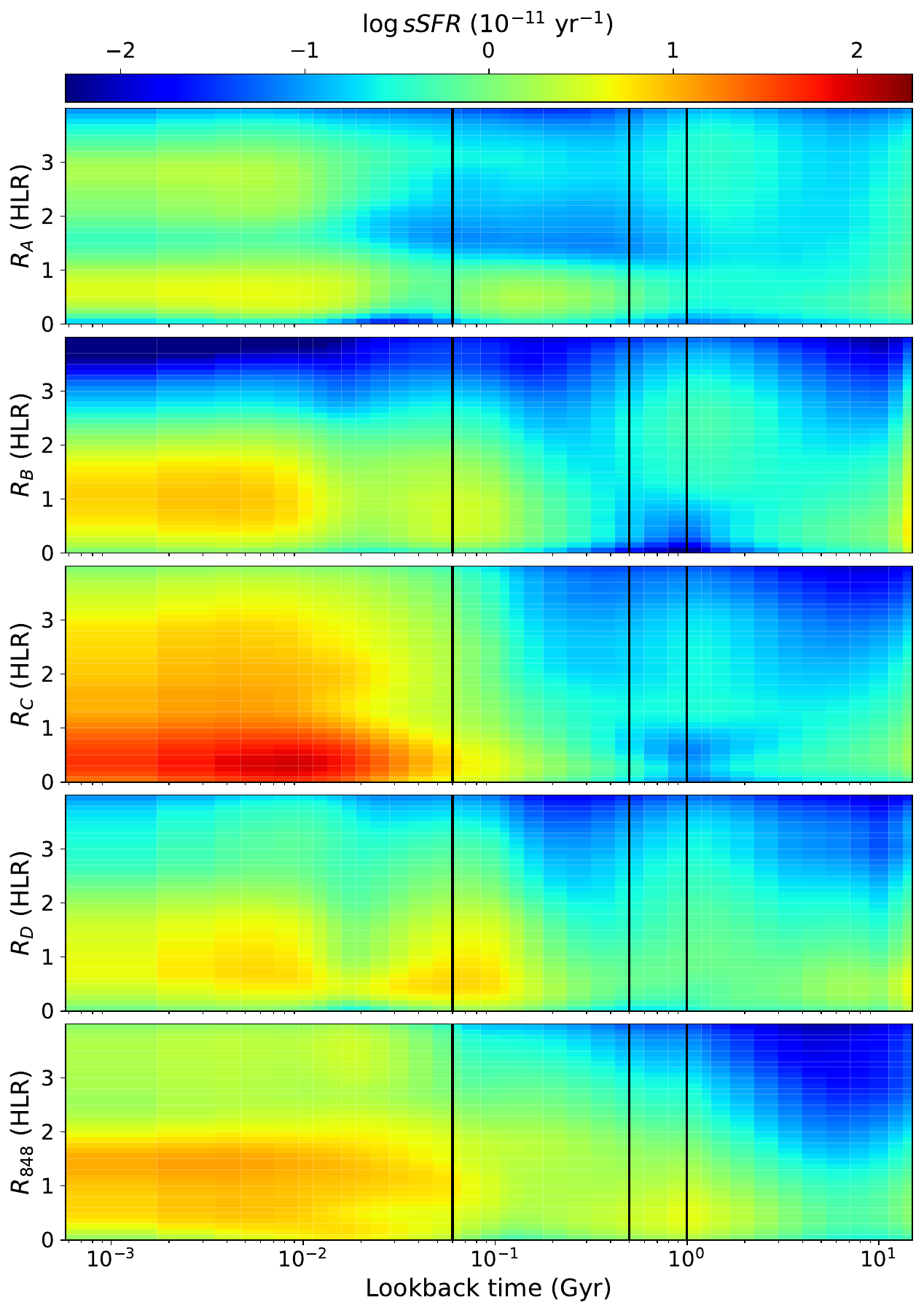}       
    \end{subfigure}
    \caption{Radial star formation history for all galaxies. The left images show the mass $M$ as a function of the radius $R$ and age while the right show the specific star formation rate $sSFR$. The data has been convolved with a Gaussian of $\approx 0.25$ dex in the lookback time axis and 0.2 HLR in the radial axis. We show vertical lines at 60 Myr, 500 Myr, and 1 Gyr for reference.}
    \label{fig:radial_convolved}
\end{figure*}

We examine the full radial distribution in mass and sSFR in Figure \ref{fig:radial_convolved}, convolved by the same Gaussian filter as Figure \ref{fig:hlr_sfh}. The centre of the first peak varies following the stellar mass of the galaxies, ranging from 2 Gyr (HCG 16a) to 1 Gyr (HCG 16d). In NGC 848, this feature is not well defined, as the sSFR does not decline after its initial rise. The 60 Myr peak is not detected in all galaxies, and in those that is there are clear differences in intensity and duration. This could indicate that while the first peak is caused by the normal evolution of the galaxies related to their mass, the second peak results from the interaction among galaxies.

\cite{osullivan2014} estimated the SFH of HCG 16b and 16c using MILES and BC03 models applied to a single SDSS-DR10 fibre (2 arcsec), observing the centre of the galaxies. For HCG 16b, they found a low-level (1-3 M$_\odot$/yr) SF component starting $\sim 300$ Myr ago (MILES) or $\sim 50$ Myr ago (BC03). For HCG 16c both models found bursts at similar epochs, but with higher SFRs: 15-50 M$_\odot$/yr for MILES and 50-80 M$_\odot$/yr for BC03.

\begin{figure}[!ht]
    \centering
    \includegraphics[width=0.50\textwidth]{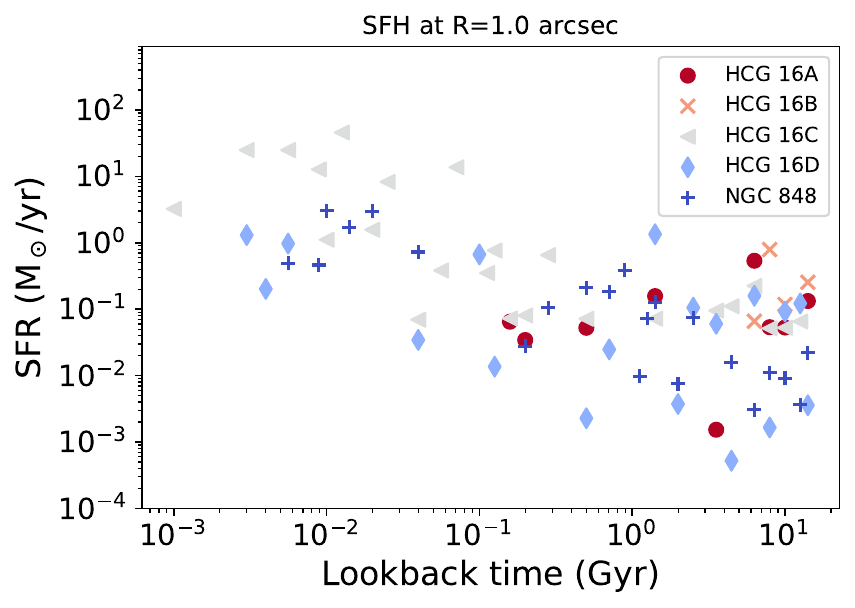}
    \caption{Star formation rate (SFR) at different epochs within 1 arcsecond radius around the nucleus, for the 5 members of HCG 16. Star formation is not detected in several epochs for the most massive galaxies. Galaxies are ordered from most massive (red) to least massive (blue), similar to Figure \ref{fig:acc_mass}.}
    \label{fig:sfh_r0.2}
\end{figure}

\begin{figure*}[!ht]
    \sidecaption
    \includegraphics[width=0.7\textwidth]{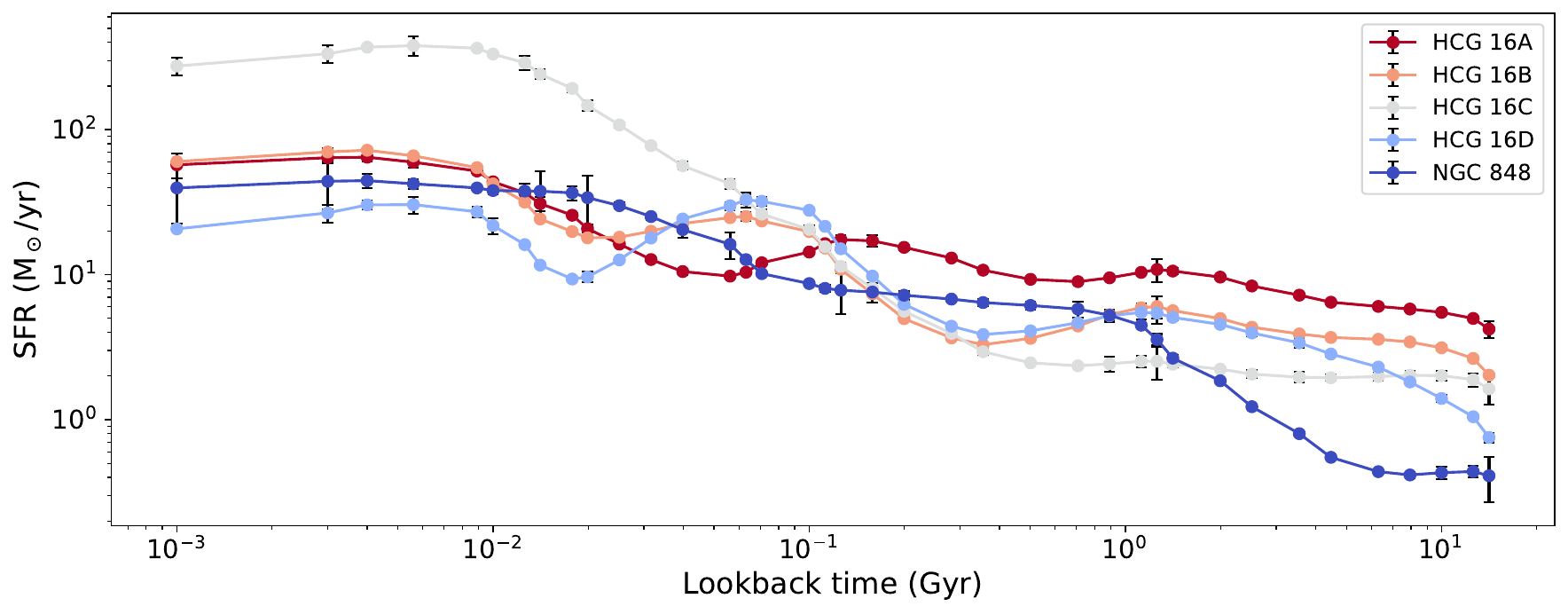}
    \caption{Integrated star formation history of HCG 16 galaxies. The data has been convolved with a Gaussian of $\approx 0.25$ dex in the lookback time axis. Galaxies are ordered from most massive (red) to least massive (blue), similar to Figure \ref{fig:acc_mass}. The error bars do not account for the full uncertainties of the stellar population synthesis.}
    \label{fig:int_sfh_conv}
\end{figure*}

Our results show little evidence of significant SF within the inner $\sim 1$–2 arcsec at most epochs, except in HCG 16c (Figure \ref{fig:sfh_r0.2}). However, at larger radii, we find consistent features: enhanced SFR around 300-400 Myr ago across all galaxies and a burst 50-60 Myr ago in HCG 16b, 16c, and 16d, consistent with a combination of the MILES and BC03 solutions. Integrated SFRs (Figure \ref{fig:int_sfh_conv}) at ages $< 60$ Myr reach from a few tens to several hundred M$\odot$ yr$^{-1}$, confirming the starburst nature of some systems. In the centre of HCG 16c, we recover SFRs consistent with the BC03 burst at $\sim 50$ Myr, but we do not detect the extended 300 Myr component, which remains at $\sim 1$ M$\odot$ yr$^{-1}$. 

\subsection{Morphology and kinematics}\label{morpho_main}

We analysed the morphology and kinematics of the galaxies by overlying \HI\ moment 0 and moment 1 contours on H$\alpha$ flux and velocity maps (figures in appendix \ref{HIHa}). We also generated interactive 3D visualisations and a movie\footnote{3D visualisations are available in \url{https://amiga.iaa.csic.es/hcg16-meerkat-muse-3d-menu/} and the movie is available online as an additional file.} of the \HI\ and H$\alpha$ datacubes with ViSL3D \citep{visl3d2025}, masked using iDaVIE \citep{idavie2024}. The main findings are discussed below, while more details can be found in Appendix \ref{morpho_app}.

Overall, the galaxies in HCG 16 show a wide range of morphologies and kinematics, with the \HI\ distribution in general more perturbed than the ionized gas. HCG 16a presents a regular velocity field and disk morphology (Figures \ref{fig:flux_coldens} and \ref{fig:mom1_mom1}), with a double ring structure in H$\alpha$ and regular resonance signals (Appendix \ref{resonance}). However, it shows an \HI\ absorption hole in the centre, a large bubble (Appendix \ref{app:bubble}), and a blueshifted tail outside the disk originated in HCG 16b \citep{silchenko2025}. HCG 16b only presents a small amount of neutral gas in the outskirts of the galaxy, and although it shows a weak and disturbed H$\alpha$ emission (Figure \ref{fig:flux_coldens}), rotation is recognised.

HCG 16c and 16d are the most disturbed galaxies. Both host strong outflows detected in H$\alpha$, while their \HI\ distributions are highly perturbed, with large absorption holes, clumps, and velocity fields that differ from the ionized gas. In HCG 16c, \HI\ traces the bridge toward HCG 16d and part of the NW tail, whereas in HCG 16d the disturbed kinematics make it difficult to distinguish between the effects of outflows and galaxy interactions. NGC 848 is the most regular galaxy in the group. Apart from outer extensions likely related to its passage through the group core and the associated \HI\ tail, its morphology and velocity field remain largely undisturbed.

\subsection{Global scenario}

The galaxies in HCG 16 show clear signatures of past interactions, such as tidal features and enhanced SF. While it is evident that interactions have played a significant role in the evolution of the group, the sequence of those encounters and their respective effects remain uncertain. Previous studies propose two scenarios for the starbursts in HCG 16c and 16d. \cite{rich2010} and \cite{vogt2013} argue that the most likely cause is their interaction with each other, while \cite{osullivan2014} and \cite{Jones2019} propose that they were triggered by the passage of NGC 848 through the group core. Simulations of encounters of galaxy pairs show that strong, sustained starbursts are typically not the outcome of interactions \citep{diMatteo2007,dimatteo2008,li2025}. Even when induced by a close passage of a galaxy, \cite{diMatteo2007,dimatteo2008} find that such bursts are usually short-lived (<150 Myr) and very rarely last up to 500 Myr. This makes the flyby scenario plausible but not strongly favoured. Moreover, \cite{vogt2013} estimate wind ages of $\sim 50$ Myr in HCG 16c, which is difficult to reconcile with a trigger event $\sim 500$ Myr ago, despite its consistency with the age of A-type populations seen in its disk. \cite{Jones2019} suggest that the flyby initiated SF that was sustained by subsequent interactions, although HCG 16a and 16b seem strongly interacting without enhanced SF.

We observe an increase in SF at $\sim400$--500 Myr (Figure \ref{fig:radial_convolved}), consistent with the proposed time of the NGC 848 passage through the group core. In HCG 16b and 16d, we find a maximum at $50$--$60$ Myr coinciding with the estimated starburst age of HCG 16c, where the SF continues to increase after 60 Myr until the present. The \HI\ distribution is highly perturbed, particularly around HCG 16c and 16d, while 16a and 16b are \HI\ deficient. These results suggest that the passage of NGC 848 contributed to a global enhancement of SF and redistribution of \HI\ in the group. However, it occurred too early to have triggered the current starbursts in HCG 16c and 16d on its own, implying that subsequent interactions were required.

HCG 16b and 16c exhibit similar SFH, diverging only in the last 50 Myr, including a pronounced increase in SF that might suggest an interaction between the two galaxies approximately 1 Gyr ago (Figure \ref{fig:hlr_sfh}). Given their projected distance of 70 kpc, a relative velocity of $\sim70$ km/s would be required, which is consistent with the typical relative velocities observed in HCGs \citep{Hickson1992}. The interaction is coherent with their disturbed morphologies and the H$\alpha$ tails observed in 16b, which are not connected to HCG 16a despite their spatial overlap. In contrast, HCG 16a shows a relatively flat SFH and an unperturbed structure --- as indicated by its Lindblad resonances (Appendix \ref{resonance}) ---, consistent with a very weak interaction with HCG 16b. Nevertheless, this evolutionary sequence remains speculative and requires further research.

We propose the following evolutionary scenario. Early interactions among core galaxies disturbed the \HI\ reservoir, forming an extended, loose envelope. Around 500 Myr ago, NGC 848 passed through the group core and stripped part of the envelope, creating the large \HI\ tail and leaving HCG 16a and 16b \HI\ deficient. Subsequent interactions among core galaxies enhanced SF in HCG 16c and 16d, but not in HCG 16a and 16b. Therefore, we find that the evolution of HCG 16 is more consistent with multiple encounters among its members than with a single triggering event.

\section{Summary}

In this work, we combined MUSE/VLT and MeerKAT observations to study the evolution, morphology, and kinematics of the galaxies in HCG 16, to understand the role of interactions within the group. We focused on the \HI\ and H$\alpha$ emission lines and on the SPS to analyse the connection between SF and neutral hydrogen. Our main findings are summarised as follows:

\begin{itemize}
    \item The galaxies in HCG 16 experienced two major episodes of enhanced SF at late evolutionary stages: the first 1--2 Gyr ago, not simultaneously in all galaxies and more clearly identified in their outer regions; the second began 400--500 Myr ago, coinciding with the estimated passage of NGC 848 through the group core, and peaked 50--60 Myr ago in HCG 16b and 16d while it has continued increasing to the present day in HCG 16c.
    \item The temporal coincidence suggests a connection between the passage of NGC 848 and the recent SF increase; however, the $\sim350$ Myr delay argues against a direct causal relationship. Instead, the passage may have triggered an initial episode later amplified by additional group interactions.
    \item HCG 16b exhibits strong morphological and kinematic perturbations whereas nearby HCG 16a appears regular, suggesting they did not interact strongly.
    \item The similar SFH of HCG 16b and 16c, together with their disturbed morphologies and the tails associated with 16b, may indicate an interaction before the passage of NGC 848 between the two galaxies.
    \item In contrast, NGC 848 and HCG 16a, which interacted less strongly than the other three galaxies, appear to follow an SFH more dependent on their mass than on interactions.
    \item Our results support a scenario in which the starbursts were driven by multiple interactions over the last few Gyr, rather than solely by the passage of NGC 848 or the HCG 16c--HCG 16d interaction. HCG 16a was mainly affected by the passage of NGC 848, which stripped most of its \HI, while leaving its disk largely unperturbed.
\end{itemize}

\section*{Data availability}
The data and code necessary to reproduce the results presented in this work are available in \url{https://github.com/ixakalabadie/HCG16-MUSE-MeerKAT}.

\begin{acknowledgements}
The authors acknowledge financial support from the grant CEX2021-001131-S funded by MICIU/AEI/ 10.13039/501100011033, from the grant PID2021-123930OB-C21 and PID2024-155817OB-I00 funded by MICIU/AEI/ and by ERDF/EU.
The authors also acknowledge the Spanish Prototype of an SRC (SPSRC) service and support funded by the Ministerio de Ciencia, Innovación y Universidades (MICIU), by the Junta de Andalucía, by the European Regional Development Fund (ERDF) and by the European Union NextGenerationEU/PRTR. The SPSRC acknowledges financial support from the AEI through the "Center of Excellence Severo Ochoa" award to the IAA-CSIC (SEV-2017-0709).
IL acknowledges financial support from the grant PRE2021-100660 funded by MICIU/AEI and by ESF+.
The authors used Claude (Sonnet 5.5, Anthropic) and ChatGPT (GPT-6, OpenAI) to assist with software development and language editing.
\end{acknowledgements}

\bibliographystyle{aa}
\bibliography{bibliography}

\begin{appendix}

\section{Morphology and kinematics}\label{morpho_app}

In this appendix, we expand the description of the morphology and kinematics given in Section \ref{morpho_main}. In particular, we analyse a bubble found in HCG 16a and determine the locations of its Lindblad resonances (LR) and corotation radius (CR). The features described in this section are observed in Figures \ref{fig:flux_coldens} and \ref{fig:mom1_mom1}, in the 3D visualisations mentioned in Section \ref{morpho_main}, and in the movie available online as an additional file.

\subsection{HCG 16a}

HCG 16a presents a double ring structure in H$\alpha$ connected by spiral arms. The inner ring contains a faint nucleus coincident with an \HI\ absorption hole of approximately one beam. High \HI\ column density clumps overlap with the outer H$\alpha$ ring except in the N, where another small \HI\ hole is found over a faint, blueshifted, H$\alpha$ region associated with a large bubble (Appendix \ref{app:bubble}). The velocity field is mostly regular in both wavelengths, except for the bubble and a similarly blueshifted H$\alpha$ tail extending to the north of the disk that appears to connect to the bubble. The 3D visualisation presents a clearer view of the features (Figure \ref{fig:bubble}) and suggests that the northern tail and the bubble are distinct structures, although the bubble H$\alpha$ emission partially blends with the disk, [\ion{N}{II}], and the northern tail, making it difficult to separate. \cite{Jones2019} reported hints of a warp in the outskirts, but this is not seen in our H$\alpha$ data or by \cite{mendesDeOliveira1998}, and we do not detect \HI\ there. \cite{silchenko2025} concluded that the blueshifted tail belongs to HCG 16b based on CALIFA \citep{CALIFA2016} data. In our observations, HCG 16b is highly disturbed and lies at a velocity similar to the tail, whereas HCG 16a is $\sim200$ km/s away and remains largely undisturbed (Appendix \ref{resonance}), supporting the association of the tail with HCG 16b. A faint signature of the bubble is also visible in the images of \cite{silchenko2025}, although they do not discuss it.

\subsection{HCG 16b}

HCG 16b is the most \HI\ deficient galaxy in the group, with only a small clump SW of the optical centre detected at high resolution. H$\alpha$ emission is also weak and disturbed, although the underlying rotation is still recognisable. A tail seems to extend toward the east at approximately the rotation velocity, matching the velocity of the \HI\ clump. The 3D visualisation reveals additional faint \HI\ near the galaxy centre and in the east.

\subsection{HCG 16c}

HCG 16c contains an \HI\ absorption hole larger than one beam, coincident with the H$\alpha$ centre. Two major \HI\ clumps are detected to the NW and E, while the central hole extends into an underdensity toward the SW. H$\alpha$ reveals two strongly redshifted outflows (N and S--SW), consistent with shock-driven winds \citep{vogt2013}; the southern outflow approximately coincides with the \HI\ underdensity, and an \HI\ valley in the NE may be associated with the northern outflow. The central H$\alpha$ velocity field shows regular rotation from E to W, whereas the disturbed \HI\ velocity field indicates rotation along a different axis (SE--NW). The \HI\ distribution also traces the bridge toward HCG 16d in the SE and part of the NW tail.

\subsection{HCG 16d}

HCG 16d exhibits the most disturbed \HI\ morphology. A large absorption hole covers several beams in the central-eastern region of the galaxy and is surrounded by numerous clumps. H$\alpha$ shows the two strongly redshifted outflows discussed by \cite{rich2010} to the N and S, together with a rotation pattern from E to W. The southern outflow appears to have displaced neutral gas creating an \HI\ underdensity in the SE, where the H$\alpha$ emission is strongest. The \HI\ velocity field suggests rotation roughly perpendicular to that of H$\alpha$, although absorption and the surrounding clumps complicate the interpretation, with some redshifted gas also present in the N.

\subsection{NGC 848}

NGC 848 is a mostly regular barred galaxy, except for extensions toward the SE, SW, and NW that are probably related to its passage through the group core and linked to the large \HI\ tail reported by \cite{VerdesMontenegro2001}. These features are visible in both wavelengths. The velocity field is otherwise regular, except for higher velocities in the centre. The \HI\ distribution displays enhanced density toward the disk edges and a lower-density central region.

\subsection{Bubble in HCG 16a}\label{app:bubble}

\begin{figure}[ht!]
    \centering
    \includegraphics[width=0.4\textwidth]{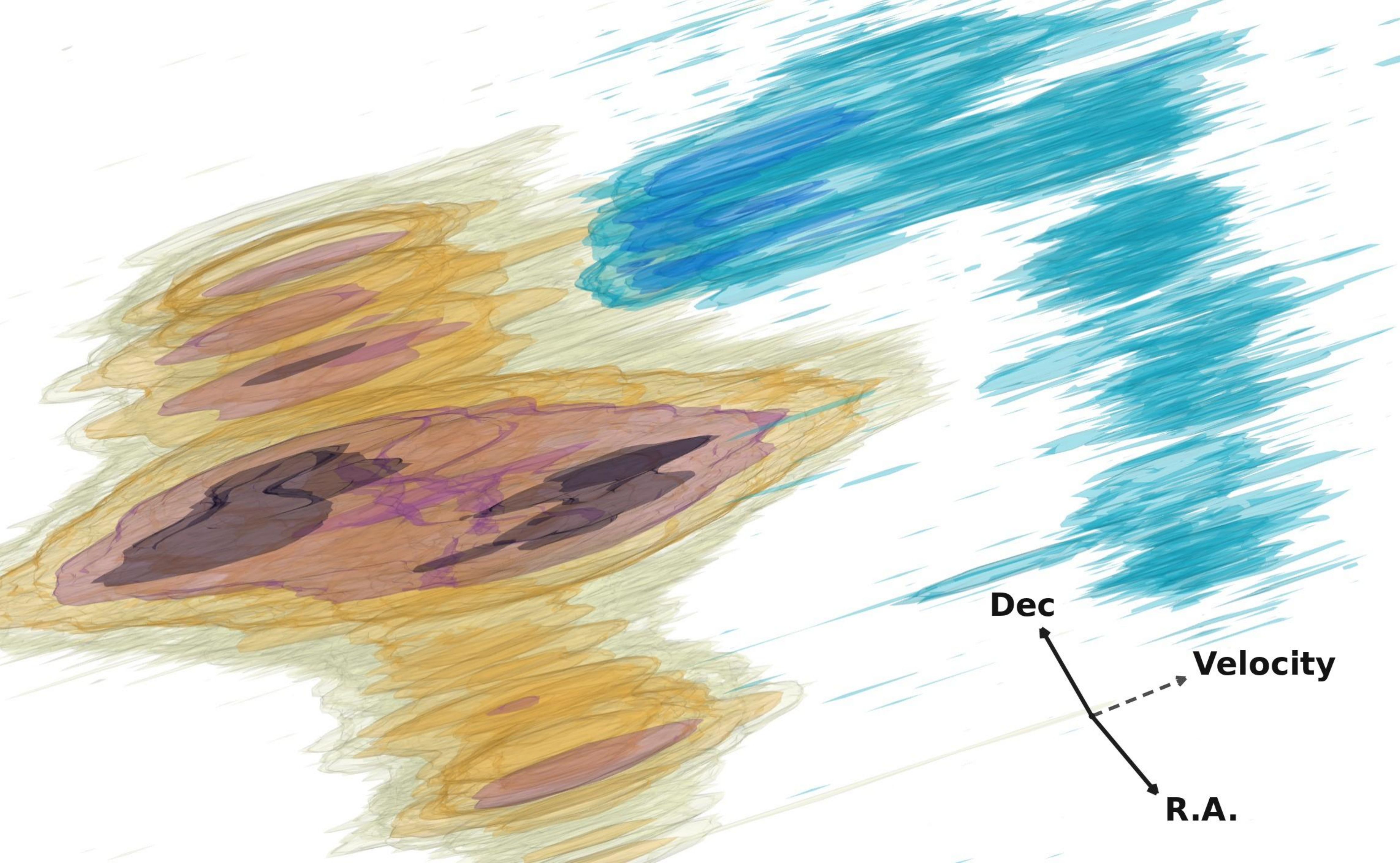}
    \caption{Image of the 3D visualisation of the H$\alpha$ emission in HCG 16a. The bubble and the tail separated from the disk are shown in different colour. The disk is shown by iso-surfaces of 5, 200, and 1000 $rms$, and the bubble and tail by iso-surfaces of 3, 10, and 20 $rms$. The $rms$ is $5.6\times10^{-20}$ erg/s/cm$^{2}$/\AA. The associated movie is available online.}
    \label{fig:bubble}
\end{figure}

The H$\alpha$ image of HCG 16a in Figure \ref{fig:mom1_mom1} reveals a blueshifted region in the northern part of the galaxy, where the regular rotational pattern is redshifted. An \HI\ hole is detected at approximately the same position in Figure \ref{fig:flux_coldens}, suggesting a relation between the ionised and neutral gas features. This structure is also visible in the 3D visualisations (e.g., Figure \ref{fig:bubble}). We interpret these features as a large expanding bubble, likely driven by stellar feedback from a star-forming region. This interpretation is also supported by the location of the bubble nearby the outer ring and by enhanced H$\alpha$ emission detected in its surroundings.

\begin{figure}
    \centering
    \begin{subfigure}{0.4\textwidth}
        \includegraphics[width=\textwidth]{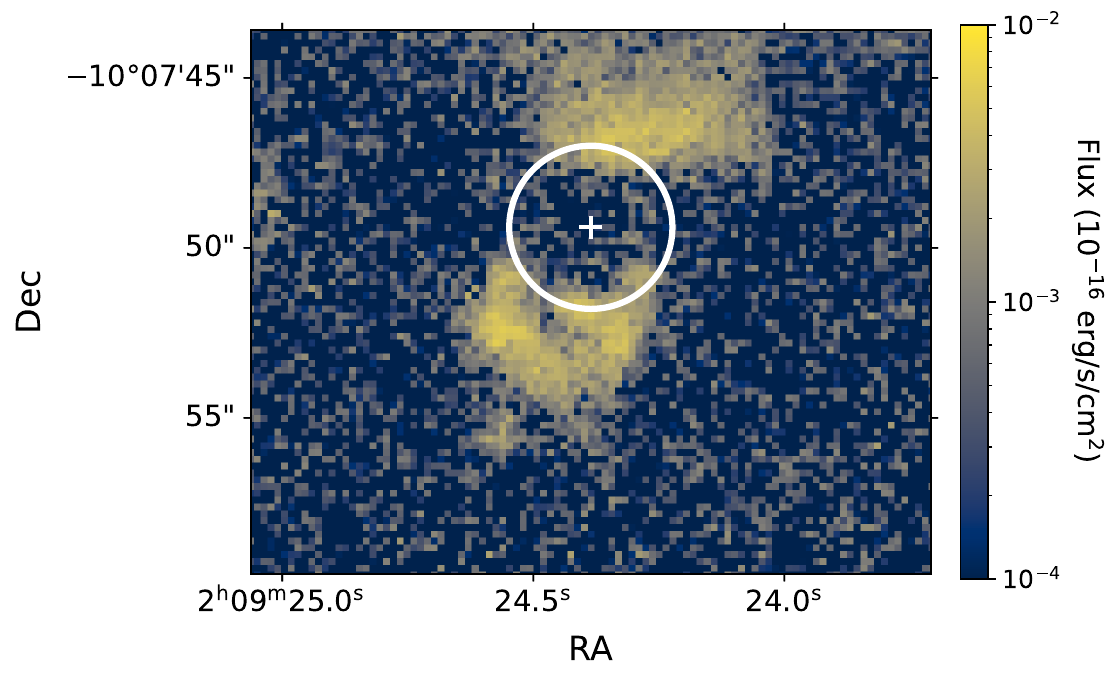}
    \end{subfigure}
    \hfill
    \begin{subfigure}{0.4\textwidth}
        \includegraphics[width=\textwidth]{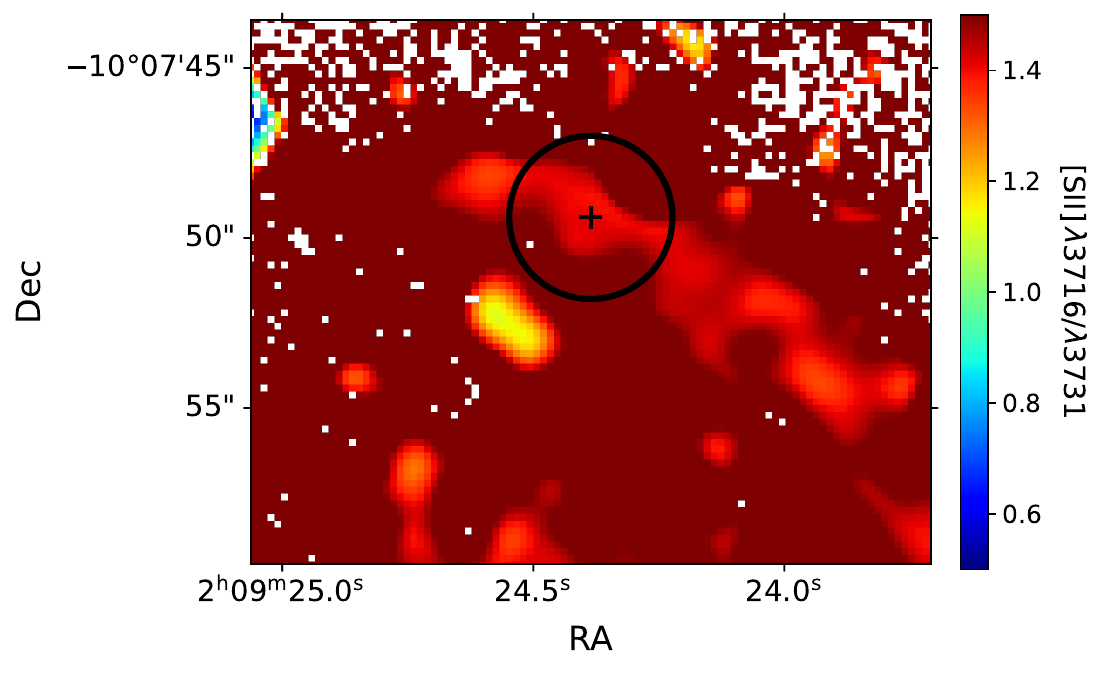}       
    \end{subfigure}
    
    \caption{H$\alpha$ flux at 3657 km/s (top) and the [\ion{S}{II}] $\lambda6716/\lambda6731$ intensity ratio (bottom) with markers at the estimated position of the bubble. The [\ion{S}{II}] data has been convolved with a $2.5''$ Gaussian kernel.}
    \label{fig:sulphur_ratio}
\end{figure}

Figure \ref{fig:sulphur_ratio} shows the H$\alpha$ emission at a velocity of 3567 km/s and the sulphur [\ion{S}{II}] $\lambda6716/\lambda6731$ intensity ratio in the region of the bubble. H$\alpha$ emission reveals shock structures produced by the stellar winds, predominantly in the declination axis, and allows us to estimate the size of the bubble. The [\ion{S}{II}] ratio suggests an elevated electron density area associated with the shocks in the south-east of the bubble, which overlaps with part of the H$\alpha$ emission in this region.

We estimated the expansion velocity, radius, and age of the bubble. Determining the expansion velocity is complicated by the overlap of the bubble with the [\ion{N}{II}] line and with the H$\alpha$ emission of the outer ring. We estimated the expansion velocity from the velocity range over which the bubble emission is detected in H$\alpha$, obtaining $v = (143\pm 57)\, \mathrm{km/s}$. To estimate the bubble radius, we extracted spectra along the declination, where the shocks are more clearly detected. The radius was then estimated from the mean separation between the two peaks associated with the expanding structure, obtaining $r = (643\pm 40)\, \mathrm{pc}$. Assuming a simple expansion scenario, we estimated the age of the bubble as $t = r/v$, obtaining $t = (4.4\pm1.8)\,\mathrm{Myr}$. These results are compatible with typical values reported for large bubbles \citep{martin1998}.

The relatively high expansion velocity may be related to the expansion of the bubble in a low-density environment. In such conditions, the interaction between the expanding gas and the surrounding medium can produce champagne flows \citep{tenorio1979}, which could contribute to the irregular morphology observed in Figure \ref{fig:sulphur_ratio}.

\subsection{Resonances in HCG 16a}\label{resonance}

We analysed the formation of structures in HCG 16a, the only one with visible and regular inner structure in the group, by estimating its CR and LR. We obtained the velocity profile from a 10 pixel (1 arcsec) wide slit along the major axis of the galaxy in the H$\alpha$ moment 1 image. Since in the N part there is a disruption on the velocity field we take just the blueshifted part of the galaxy. We also calculated the epicyclic frequency $\kappa$, which is used to determine the inner and outer LR. We used the Tremaine-Weinberg method \citep{Tremaine1984} to estimate the pattern speed $\Omega_P$ using the stellar velocity image, which is preferred to clumpy tracers like H$\alpha$ \citep{williams2021}. From the results, we have estimated the inner Lindblad resonance (ILR) and the CR.

\begin{figure}[ht!]
    \centering
    \includegraphics[width=0.4\textwidth]{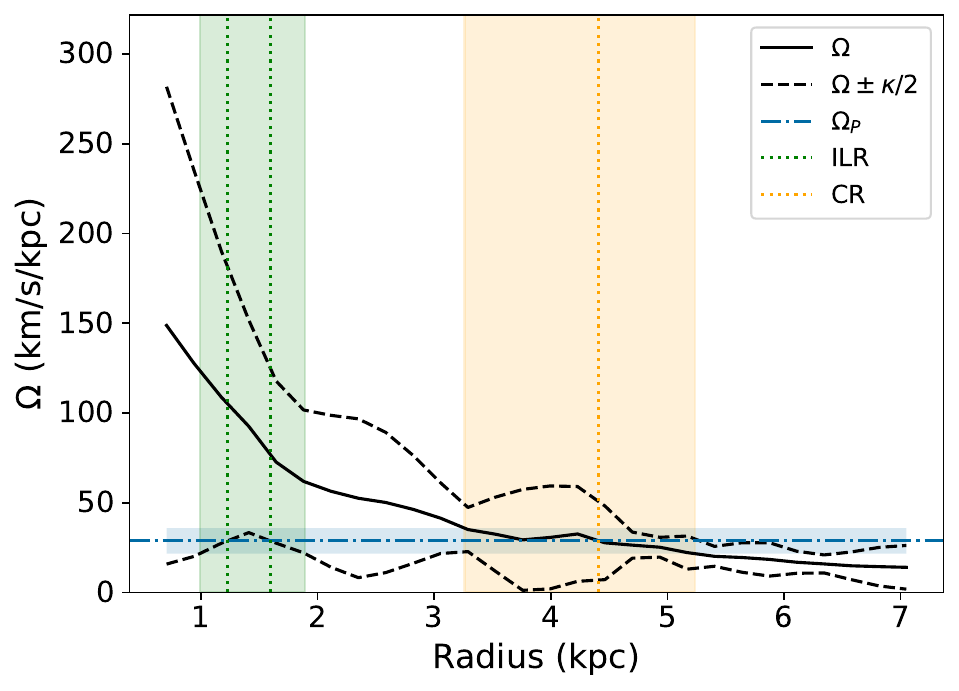}
    \caption{Angular velocity $\Omega$ against the radius at the blueshifted side of the galaxy. We also show the pattern speed $\Omega_P$, the ILR, and the CR with their uncertainties.}
    \label{fig:omega}
\end{figure}

\begin{figure}
    \centering
    \includegraphics[width=0.4\textwidth]{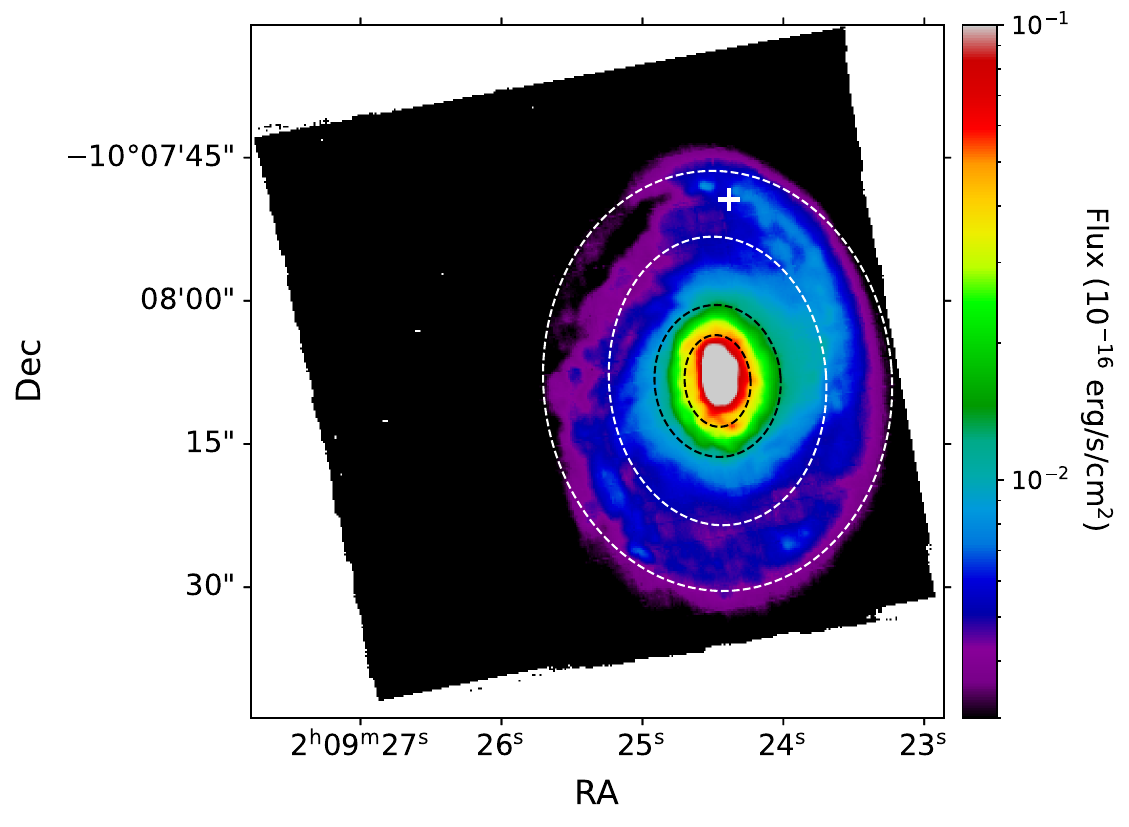}
    \caption{V-band image of HCG 16a with CR 1$\sigma$ uncertainty (white rings) and the equivalent for the ILR (black rings) overlaid, displayed to enhance the bar and disks in the galaxy. The white cross indicates the approximate position of the bubble.}
    \label{fig:vband}
\end{figure}

Figure \ref{fig:omega} shows the angular frequency $\Omega$ and $\Omega \pm \kappa/2$ at different radii and the pattern speed ($29\pm7$ km/s/kpc). The CR is found where $\Omega_P$ crosses $\Omega$, in this case at a radius of $4.3\pm1.0$ kpc. The ILR is found where $\Omega_P$ crosses $\Omega \pm \kappa/2$ and we find two radii where this happens; these could be the inner and outer ILR. However, the uncertainties are too large to properly characterise them, therefore, we have taken it as a single ILR at $1.4\pm0.5$ kpc. The resonances are represented in Figure \ref{fig:vband} over the V-band image, modified to highlight the inner structure of the galaxy. The image shows the bar as well as the two rings, which coincide with resonance radii.

The fact that we can calculate resonances in HCG 16a indicates that the galaxy is dynamically stable and the disk is evolving unperturbed. This is in contrast with the expectation in a highly interacting group, and the rest of the galaxies show the expected perturbation signals. Notably, the closest galaxy, HCG 16b, shows a very disturbed disk, indicating that HCG 16a might not have interacted as strongly with the rest of the galaxies as the rest among themselves.

\clearpage
\onecolumn
\section{Additional figures}\label{HIHa}

\begin{figure*}[!ht]
    \centering
    \begin{subfigure}{0.3\linewidth}
        \includegraphics[width=\linewidth]{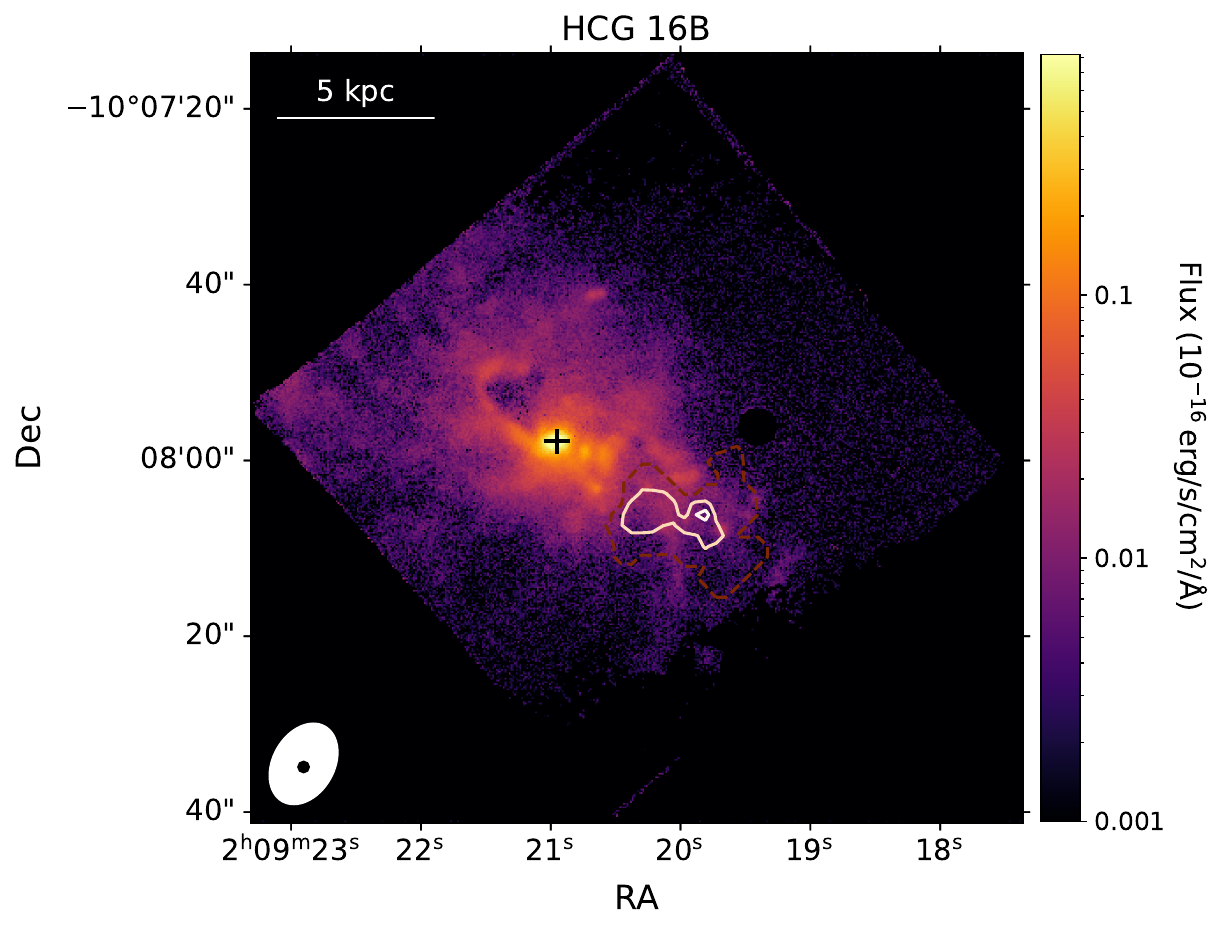}
    \end{subfigure}
    \hfill
    \begin{subfigure}{0.3\linewidth}
        \includegraphics[width=\linewidth]{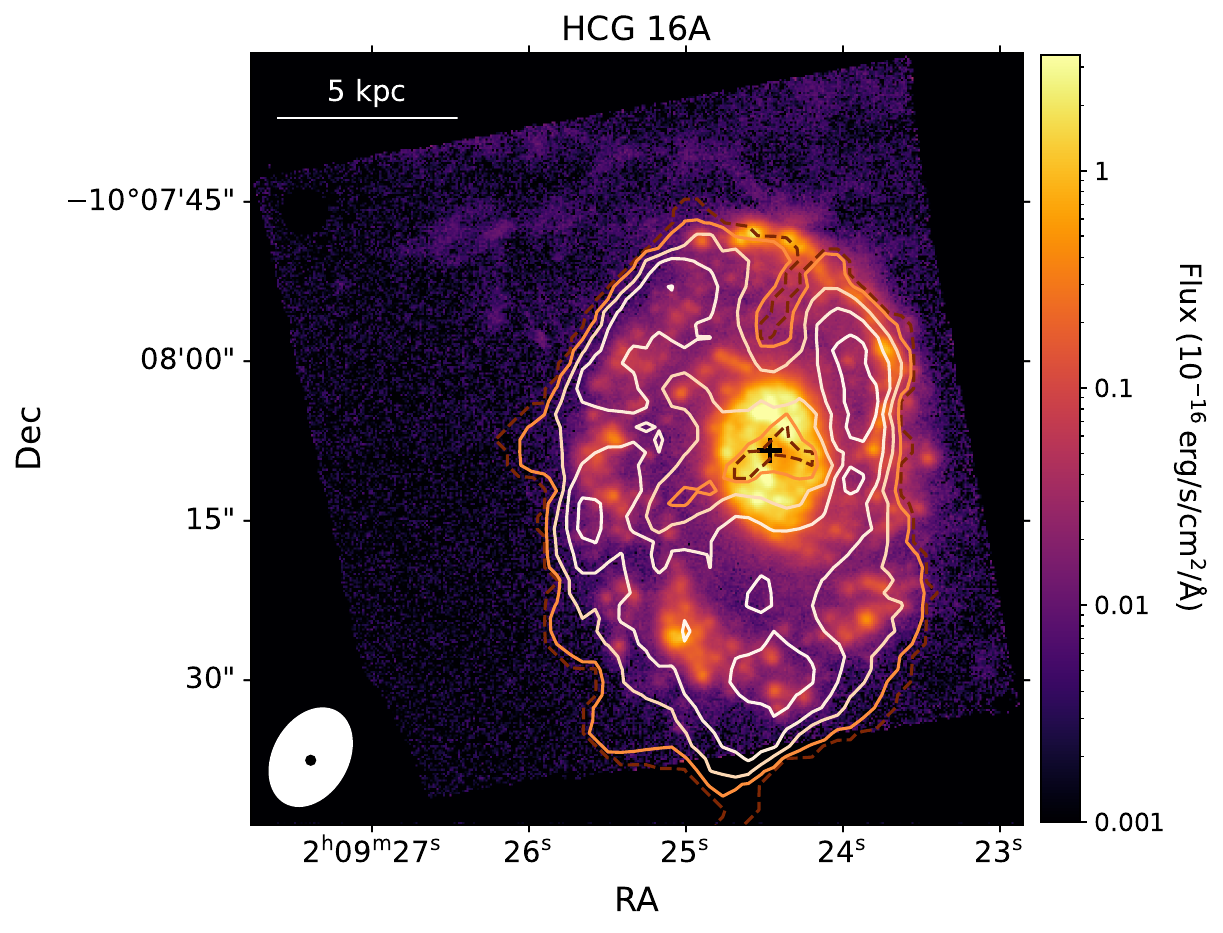}       
    \end{subfigure}
    \hfill
    \begin{subfigure}{0.3\linewidth}
        \includegraphics[width=\linewidth]{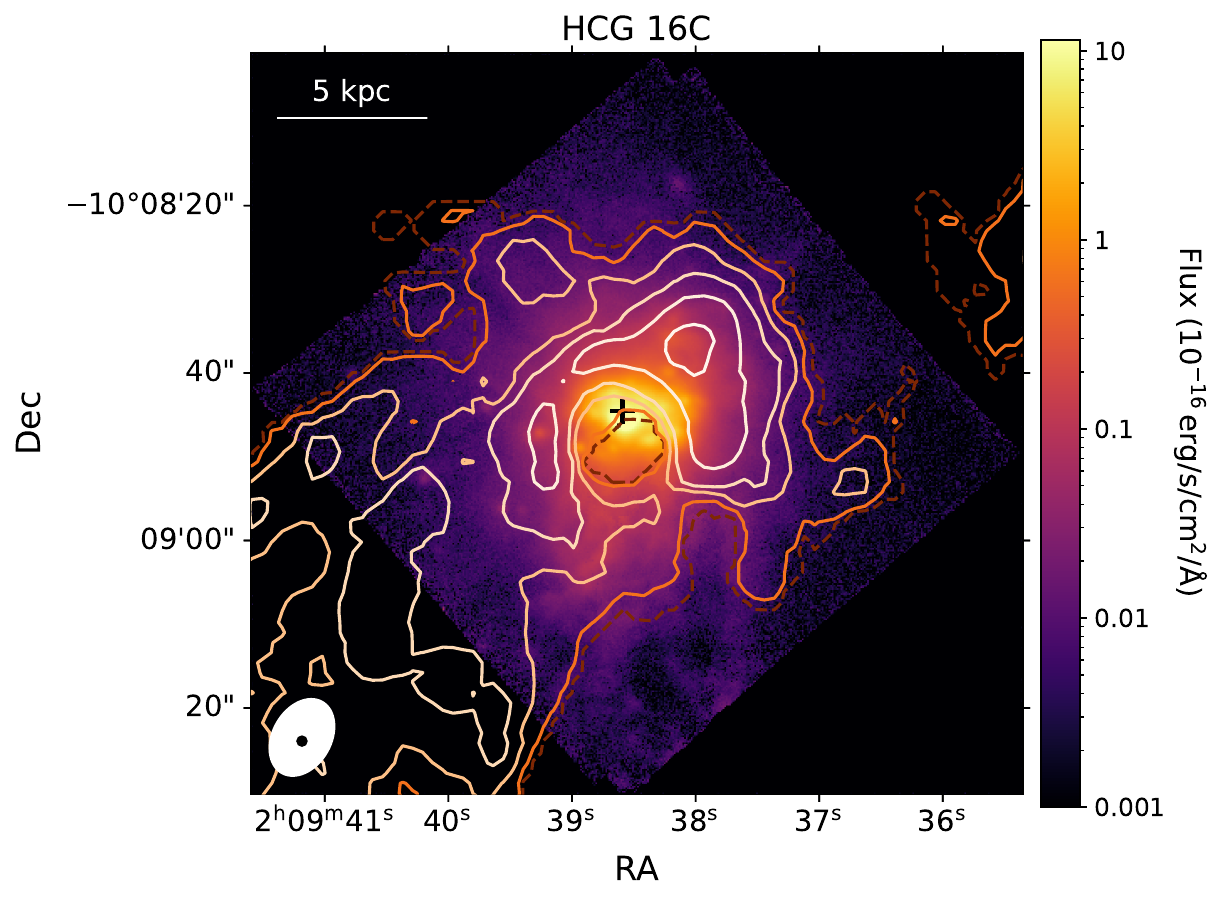}
    \end{subfigure}
    
    \makebox[\linewidth][c]{
        \begin{subfigure}{0.3\linewidth}
            \includegraphics[width=\linewidth]{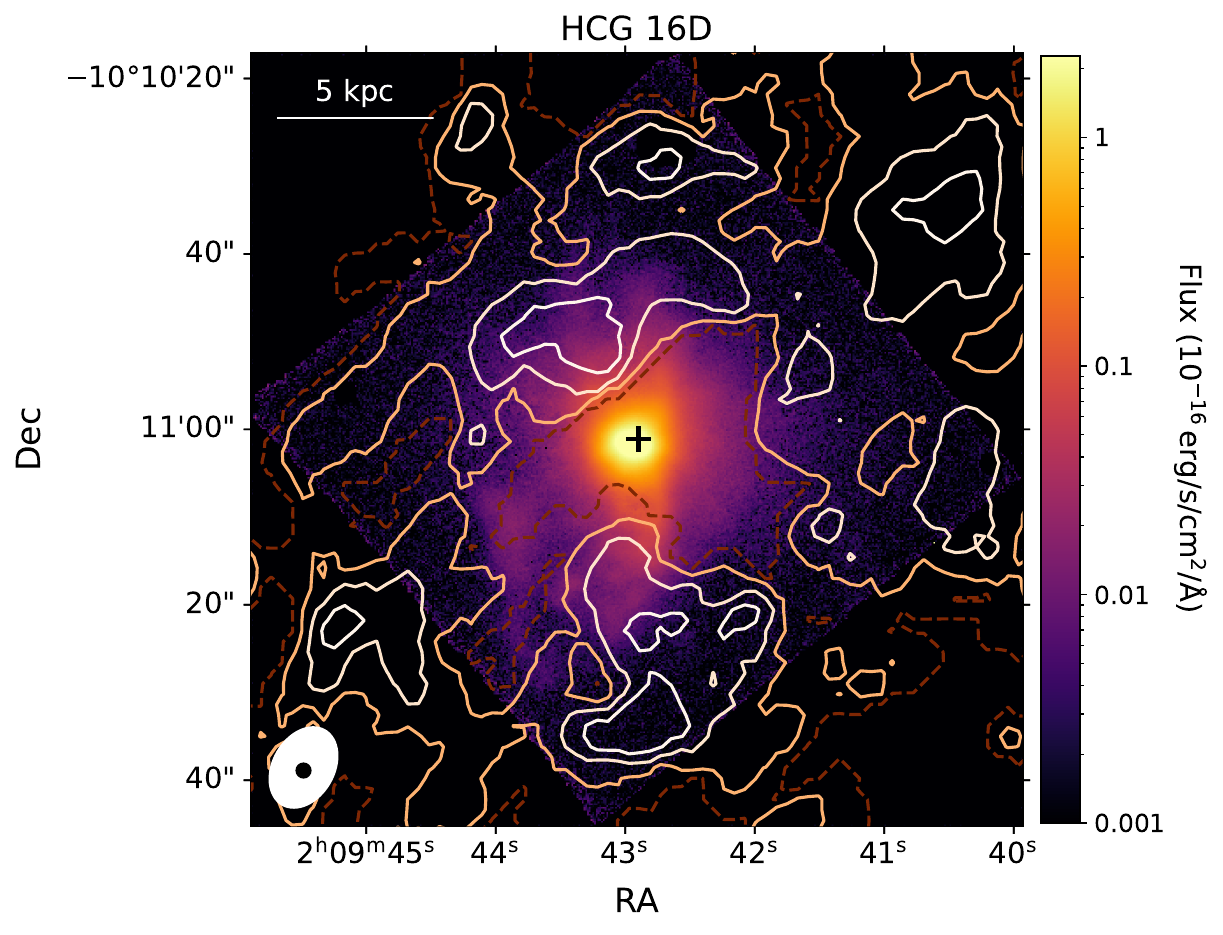}
        \end{subfigure}
        \hspace{0.02\linewidth}
        \begin{subfigure}{0.3\linewidth}
            \includegraphics[width=\linewidth]{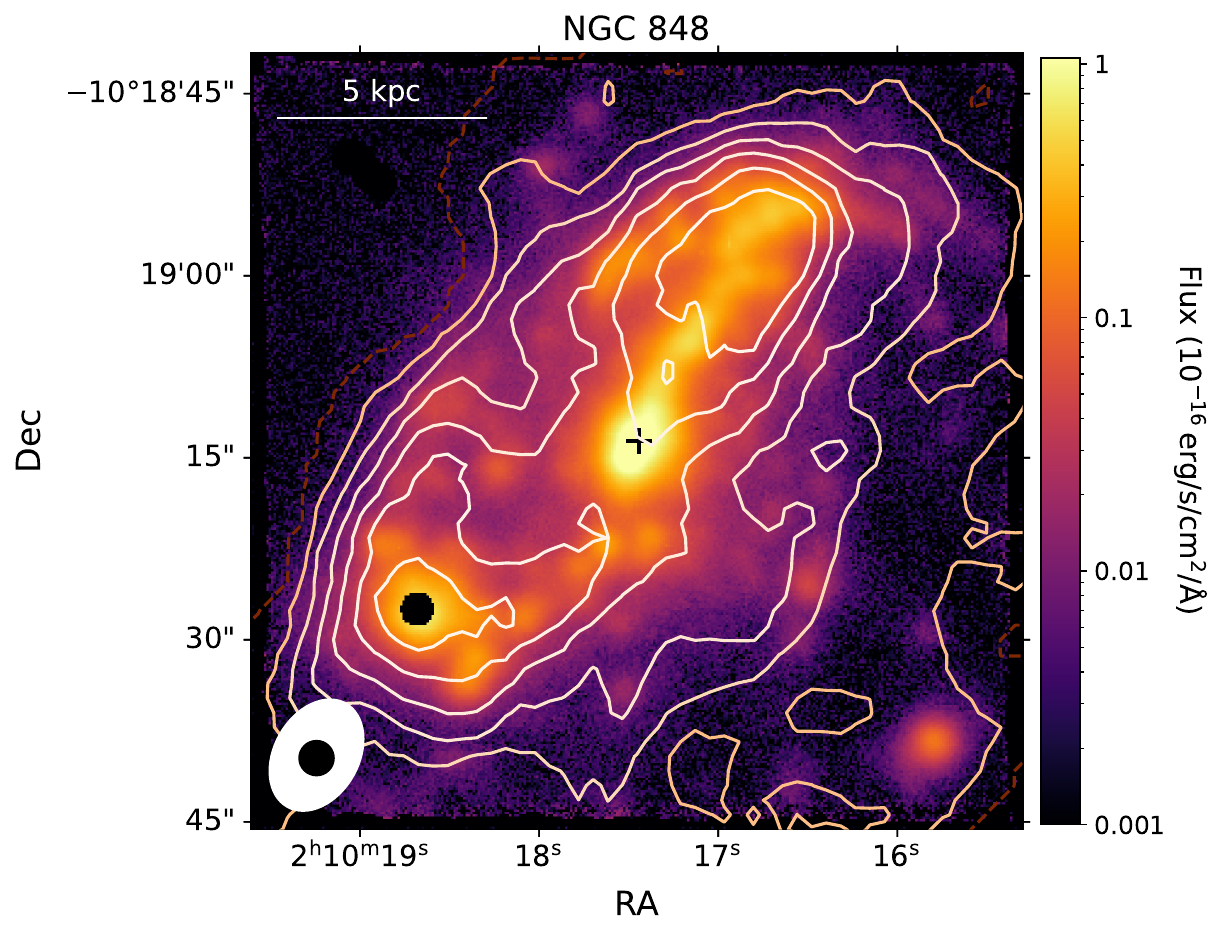}
        \end{subfigure}
    }

    \caption{HI column density contours overlaid on H$\alpha$ flux images. The black cross represents the centre of the galaxy in H$\alpha$. The white ellipse is the MeerKAT beam and the overlaid black circle is the MUSE PSF. Top-left: HCG 16b with contour levels $1\times10^{19}$, $2\times10^{20}$, $4\times10^{20}$ cm$^{-2}$. Top-centre: HCG 16a with contour levels $1\times10^{19}$, $1\times10^{20}$, $4\times10^{20}$, $7\times10^{20}$, $1\times10^{21}$ cm$^{-2}$. Top-right: HCG 16c with contour levels $1\times10^{19}$, $1\times10^{20}$, $5\times10^{20}$, $1\times10^{21}$, $2\times10^{21}$, $3\times10^{21}$ cm$^{-2}$. Bottom-left: HCG 16d with contour levels $1\times10^{19}$, $2\times10^{20}$, $6\times10^{20}$, $1.1\times10^{21}$ cm$^{-2}$. Bottom-right: NGC 848 with contour levels $1\times10^{19}$, $5\times10^{20}$, $1\times10^{21}$, $1.5\times10^{21}$, $2\times10^{21}$, $2.5\times10^{21}$, $3\times10^{21}$ cm$^{-2}$. The dashed contour ($1\times10^{19}$ cm$^{-2}$) marks the border of the mask. Notice that the central parts of HCG 16a, 16c, and 16d are masked due to absorption.}
    \label{fig:flux_coldens}
\end{figure*}

\begin{figure*}[!ht]
    \centering
    \begin{subfigure}{0.3\linewidth}
        \includegraphics[width=\linewidth]{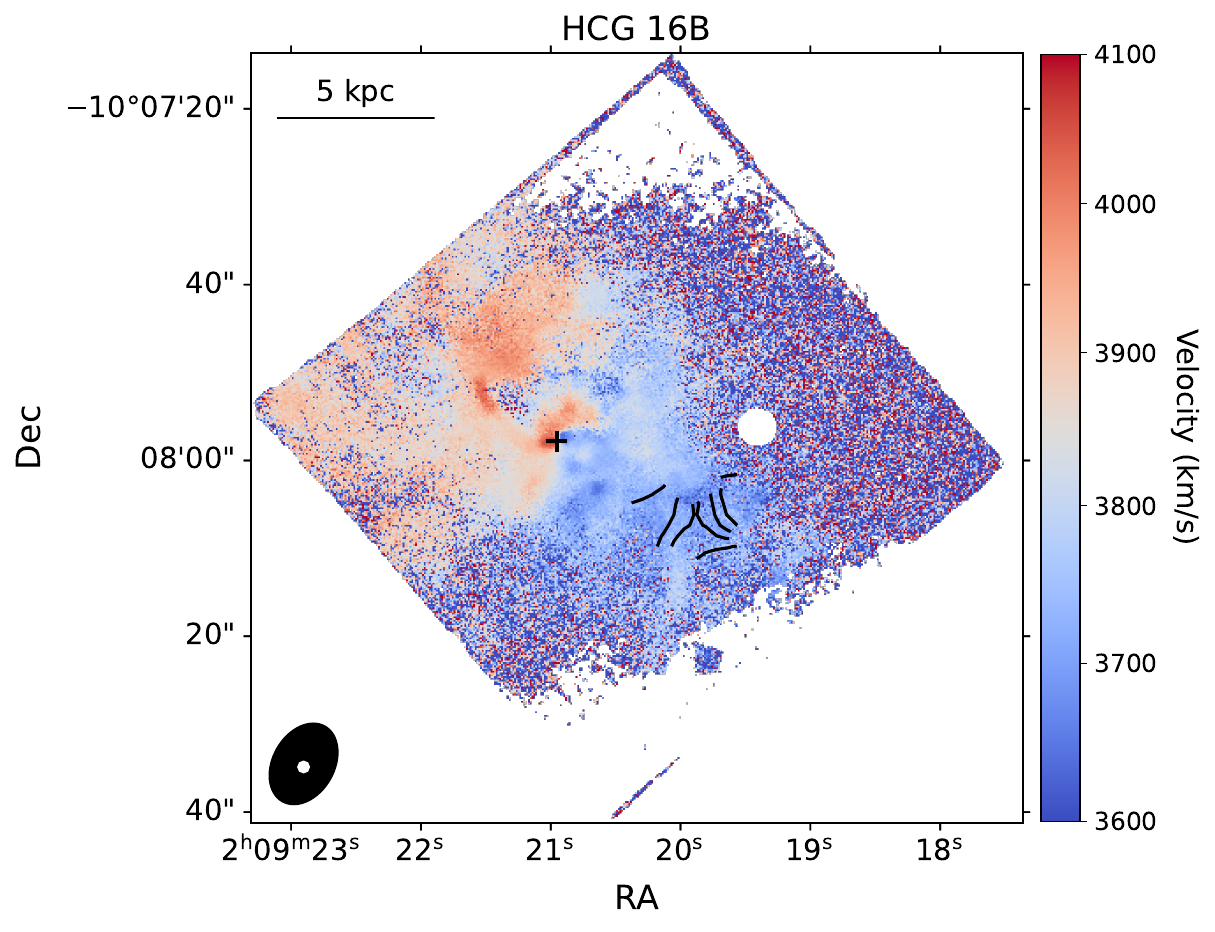}
    \end{subfigure}
    \hfill
    \begin{subfigure}{0.3\linewidth}
        \includegraphics[width=\linewidth]{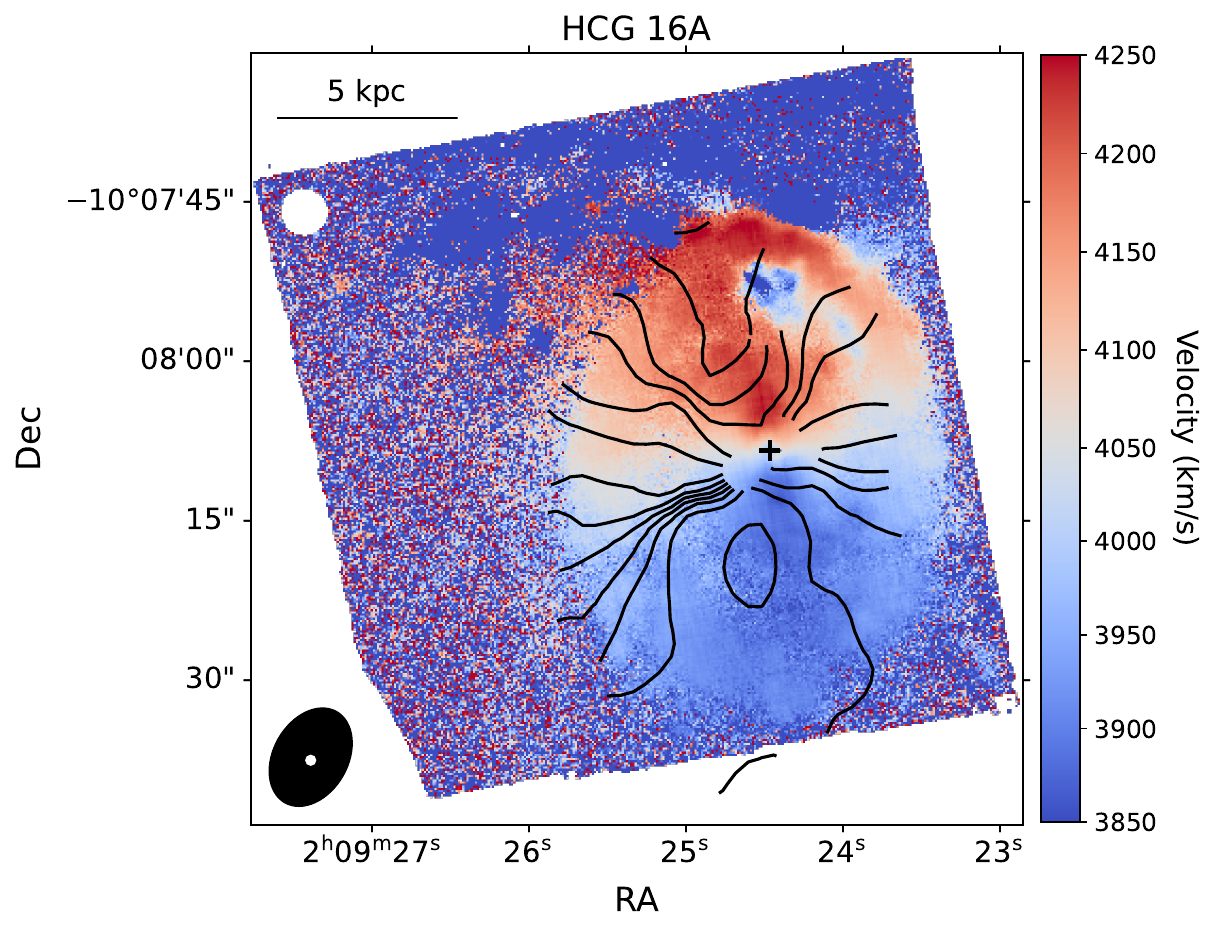}       
    \end{subfigure}
    \hfill
    \begin{subfigure}{0.3\linewidth}
        \includegraphics[width=\linewidth]{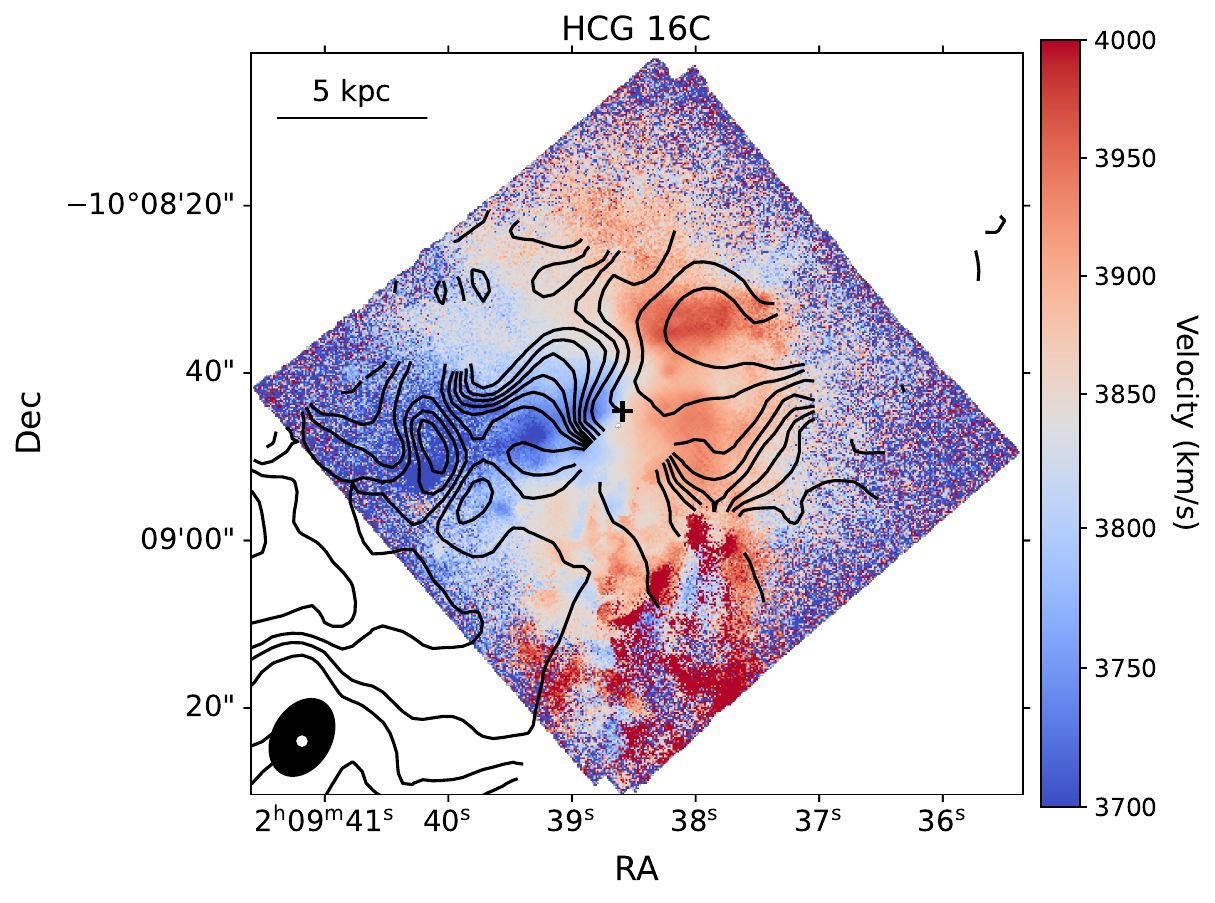}
    \end{subfigure}
    
    \makebox[\linewidth][c]{
        \begin{subfigure}{0.3\linewidth}
            \includegraphics[width=\linewidth]{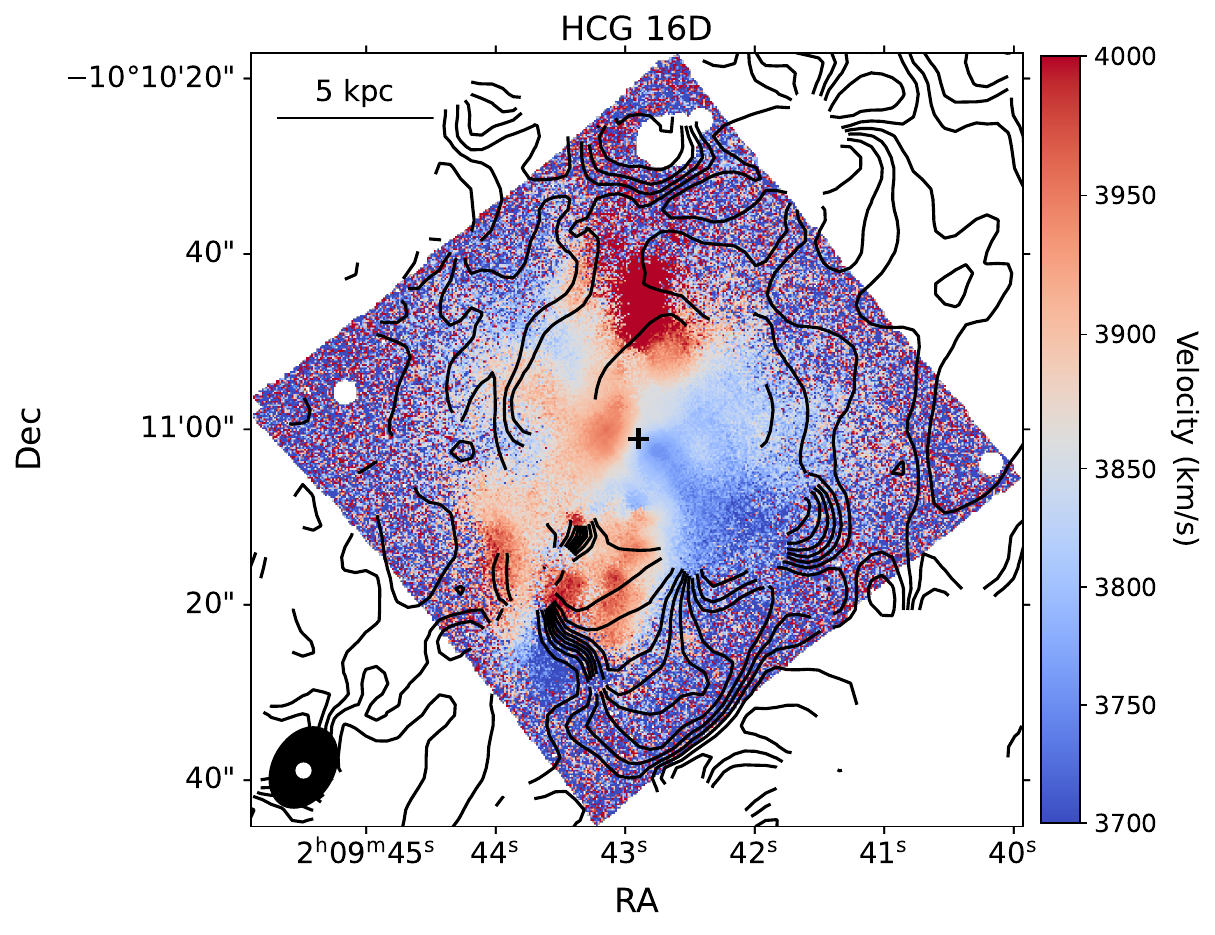}
        \end{subfigure}
        \hspace{0.02\linewidth}
        \begin{subfigure}{0.3\linewidth}
            \includegraphics[width=\linewidth]{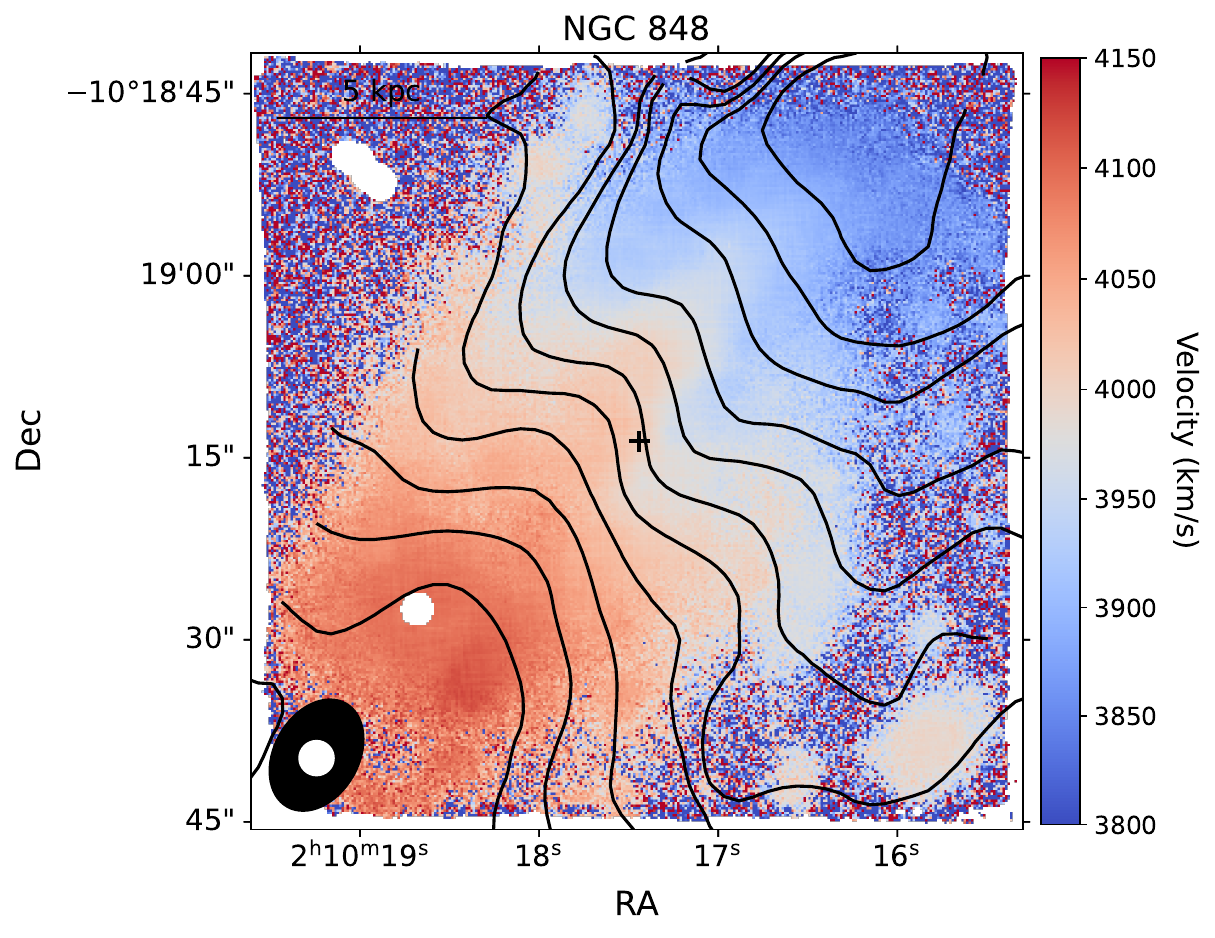}
        \end{subfigure}
    }

    \caption{HI velocity contours overlaid on H$\alpha$ velocity images. The black cross represents the centre of the galaxy in H$\alpha$. The black ellipse is the MeerKAT beam and the overlaid white circle is the MUSE PSF. Top-left: HCG 16b with contour levels from 3700 to 3780 km/s in steps of 20 km/s. Top-centre: HCG 16a with contour levels from 3860 to 4220 in steps of 30 km/s. Top-right: HCG 16c with contour levels from 3700 to 3940 in steps of 20 km/s. Bottom-left: HCG 16d with contour levels from 3760 to 4020 in steps of 20 km/s. Bottom-right: NGC 848 with contour levels 3840 to 4080 in steps of 20 km/s.}
    \label{fig:mom1_mom1}
\end{figure*}

\end{appendix}

\end{document}